\documentclass[letterpaper, 10pt, times, preprint]{elsarticle}

\usepackage[english]{babel}
\usepackage{graphicx,subcaption} 
\usepackage{multirow}
\usepackage[table,xcdraw]{xcolor}
\usepackage{array}

\usepackage{float}

\usepackage[letterpaper,top=2cm,bottom=2cm,left=3cm,right=3cm,marginparwidth=1.75cm]{geometry}

\usepackage{amsmath}
\usepackage{amssymb}

\usepackage[numbers]{natbib}

\usepackage{graphicx}
\usepackage[colorlinks=true, allcolors=blue]{hyperref}

\begin{document}

\begin{frontmatter}
\title{Hierarchical Gaussian Process-Based Bayesian Optimization for Materials Discovery in High Entropy Alloy Spaces}
\author{Sk Md Ahnaf Akif Alvi$^{a,c}$} 
\corref{mycorrespondingauthor}
\ead{ahnafalvi@tamu.edu}

\author{Jan Janssen$^{b}$}
\author{Danial Khatamsaz$^{a}$}
\author{Danny Perez$^{c}$}
\author{Douglas Allaire$^{d}$}
\author{Raymundo Arróyave$^{a,d,e}$}
    
\address{$^a$Department of Materials Science and Engineering, Texas A\&M University, College Station, TX, USA 77843}
\address{$^b$ Max-Planck-Institute for Sustainable Materials, D\"usseldorf, Germany, 40237}
\address{$^c$ Theoretical Division T-1, Los Alamos National Laboratory, Los Alamos, NM, USA 87544}
\address{$^d$ J. Mike Walker '66 Department of Mechanical Engineering, Texas A\&M University, College Station, TX, USA 77843}
\address{$^e$ Wm Michael Barnes '64 Department of Industrial and Systems Engineering, Texas A\&M University, College Station, TX, USA 77843}

\begin{abstract}
Bayesian optimization (BO) is a powerful and data-efficient method for iterative materials discovery and design, particularly valuable when prior knowledge is limited, underlying functional relationships are complex or unknown, and the cost of querying the materials space is significant. Traditional BO methodologies typically utilize conventional Gaussian Processes (cGPs) to model the relationships between material inputs and properties, as well as correlations within the input space. However, cGP-BO approaches often fall short in multi-objective optimization scenarios, where they are unable to fully exploit correlations between distinct material properties. Leveraging these correlations can significantly enhance the discovery process, as information about one property can inform and improve predictions about others. This study addresses this limitation by employing advanced kernel structures to capture and model multi-dimensional property correlations through multi-task (MTGPs) or deep Gaussian Processes (DGPs), thus accelerating the discovery process. We demonstrate the effectiveness of MTGP-BO and DGP-BO in rapidly and robustly solving complex materials design challenges that occur within the context of complex multi-objective optimization---carried out by leveraging the pyiron workflow manager---over FCC FeCrNiCoCu high entropy alloy (HEA) spaces, where traditional cGP-BO approaches fail. Furthermore, we highlight how the differential costs associated with querying various material properties can be strategically leveraged to make the materials discovery process more cost-efficient. 
\end{abstract}


\end{frontmatter}

\section{Introduction}
Accelerating technological advancements often necessitates a corresponding acceleration in the materials development cycle. Central to this overarching initiative are the Integrated Computational Materials Engineering (ICME) paradigm \cite{national2008integrated} and the Materials Genome Initiative (MGI) \cite{national2011materials}. ICME, established prior to MGI, focuses on integrating complex computational workflows to streamline the creation of new materials and processes \cite{wang2019integrated,10.1115/1.4036649,10.1007/s11837-015-1514-5,10.1007/s11837-014-1260-0,10.1051/matecconf/202032110013,10.1016/j.calphad.2015.04.002,10.1063/1.4977487}. Despite its advantages, ICME programs often face limitations due to the high computational cost of models and the challenges in creating explicit connections across different modeling steps \cite{liu2018vision}. In contrast, the MGI, launched in 2011, adopts a broader and more aspirational approach, advocating for the integration of advanced computational tools and data science to accelerate materials discovery \cite{10.1038/s41524-019-0173-4}. This initiative has facilitated the adoption of emerging technologies, such as machine learning (ML), which has become a powerful tool for addressing limitations in traditional methods, particularly in predictive modeling for materials science.

ML excels at processing vast amounts of data, enabling the identification of patterns, the prediction of material properties, and the optimization of performance \cite{10.1063/1.4946894,10.1038/s41467-018-08223-5}. By leveraging data from high-throughput simulations, fabrication processes, and characterization efforts, ML assists in identifying promising material candidates and aids in designing materials with specific functionalities \cite{10.1038/s41524-017-0006-2,10.1093/nsr/nwad125,10.55003/cast.2023.06.23.014}. When combined with physics-based models, ML can further enhance predictive insights, particularly for high-temperature alloys, thereby improving both material performance and resilience \cite{10.1038/s41524-020-00407-2,10.1038/s41598-018-36574-y,10.1002/cepa.2639,10.17762/turcomat.v9i2.13858,10.17762/turcomat.v9i2.13863}. Beyond predictive capabilities, ML is increasingly employed to explore and optimize materials spaces, framing materials discovery as a `black box' optimization problem. The objective is to optimize material properties by varying features such as chemical composition and processing conditions. Given the vast array of possible configurations and microstructures, efficient navigation of these spaces poses significant challenges. Bayesian Optimization (BO) has emerged as a key method to address this challenge \cite{frazier2016bayesian,snoek2012practical}.

BO is particularly effective for optimizing functions that are expensive to evaluate and when available information is limited. It achieves this by using: (1) a surrogate model, typically a Gaussian Process (GP), to approximate the objective function while quantifying prediction uncertainty; (2) an acquisition function that strategically balances exploration and exploitation to determine the next evaluation point; and (3) an iterative process that refines the model and acquisition function with each new experiment \cite{frazier2018tutorial, snoek2012practical}. GPs are favored in BO due to their mathematical rigor, flexibility, and ability to quantify uncertainty, making them ideal for a wide range of optimization challenges \cite{ceylan2020estimation, iyer2019data, snoek2012practical}. By effectively balancing exploration and exploitation, BO accelerates the identification of materials with desired properties while minimizing the number of required experiments \cite{iyer2019data, talapatra2018autonomous, ju2017designing, ghoreishi2018multi, khatamsaz2021efficiently, ghoreishi2019efficient, liu2020multi, janet2017resolving, ramprasad2017machine, honarmandi2021top}.

BO’s data-efficient nature makes it an invaluable tool for the design and optimization of complex materials. Although BO has been successfully applied across various materials discovery challenges, it is particularly effective in navigating the expansive compositional spaces of high-entropy alloys (HEAs). HEAs, which incorporate five or more principal metallic elements, represent a significant advancement in alloy design by offering a broad landscape for optimizing material properties \cite{10.21741/9781644902615-29}. Given the vast design space, exhaustive searches are impractical, necessitating the adoption of strategic experimentation.

BO has been instrumental in advancing HEA development by enabling the creation of alloys with enhanced mechanical and functional properties, thus overcoming the limitations of conventional alloys at relatively modest computational or experimental costs \cite{pedersen2021bayesian,torsti2024improving,halpren2024machine,liu2023machine,qian2024accelerating,sulley2024accelerating}. Its adaptability is further demonstrated through its application in exploring high-dimensional HEA design spaces using innovative approaches. These strategies include leveraging informative priors (incorporating domain expertise) \cite{vela2023data}, integrating physics-informed variants (embedding known physical laws) \cite{khatamsaz2023physics}, implementing parallelized optimization (enabling batch-wise suggestions per iteration) \cite{hastings2024interoperable}, and ensuring constraint-aware optimization (adhering to specific design constraints) \cite{khatamsaz2023bayesian}.

In practical applications, materials design always involves the consideration of multiple factors, including conflicting performance targets, design allowables, and constraints. This complexity renders single-objective BO-based materials optimization largely an academic exercise, as real-world problems require balancing trade-offs between various properties. Multi-objective BO-based materials discovery, although not as developed as single-objective BO-based materials design, has started to address these challenges \cite{10.1016/j.matdes.2018.10.014,hastings2024interoperable,zhao2018fast,gopakumar2018multi,arroyave2022perspective,biswas2023multi}. To date, most approaches to BO-based multi-objective materials design assume that each property target is independent, modeling the multi-objective response through independent GPs for each property. However, this approach is suboptimal, as it does not account for the fact that materials properties are often correlated to varying degrees, given their dependence on the same underlying arrangement of matter---examples of such correlations include the strength-ductility tradeoff or the positive correlation, within metallic systems, between density and electronic thermal conductivity. Unlike conventional GPs, Deep Gaussian Processes (DGPs) and Multi-Task Gaussian Processes (MTGPs) can learn correlations among multiple outputs using connected kernel structures, making them more efficient and effective in optimizing complex materials.

Multi-Task Gaussian Processes (MTGPs) are effective in learning correlations between related tasks, improving generalization by sharing information across tasks \cite{10.48550/arxiv.1707.08114}. MTGPs model both positive and negative correlations, which enhances prediction quality and aids in identifying outliers \cite{10.1145/2538028,10.48550/arxiv.1302.2576}. They use message-passing architectures to iteratively share latent information, efficiently transferring knowledge in correlated tasks \cite{10.18653/v1/p19-1048,dizaji2021deep}. MTGPs have been successfully applied in fields like gene expression \cite{dizaji2021deep}, sentiment analysis \cite{10.18653/v1/p19-1048}, and human motion modeling \cite{10.1109/tpami.2007.1167}, making them powerful tools for multi-task learning. Deep Gaussian Processes (DGPs) offer a hierarchical extension of GPs, combining the flexibility of deep neural networks with the uncertainty quantification of GPs \cite{10.48550/arxiv.1806.01655}. These models, equivalent to neural networks with infinitely wide hidden layers, capture complex, non-linear relationships \cite{10.48550/arxiv.2110.00568}. They have proven effective in time series forecasting \cite{10.48550/arxiv.2301.09811}, image classification \cite{10.48550/arxiv.1902.05888}, and reinforcement learning \cite{10.2991/ijcis.2018.25905189}. 

The advantages of DGP and MTGP over cGP within a BO setting remains to be demonstrated for any materials design scenario with multiple objectives, however, the potential to dramatically improve the performance of Bayesian materials discovery schemes is evident: unlike cGP-BO, which models each material property independently, MTGP-BO and DGP-BO are capable of capturing correlations between multiple material properties. This capability allows these advanced BO methods to exploit shared information across different properties, leading to more efficient exploration and optimization of complex material spaces. By leveraging these correlations, MTGP-BO and DGP-BO can reduce the number of required experiments, accelerate the discovery process, and achieve more accurate predictions in multi-objective optimization tasks.

One promising application of these advanced Bayesian Optimization (BO) methods lies in the discovery of alloys that must simultaneously satisfy multiple, interrelated performance criteria. A prime example of such a discovery problem is the identification of alloys that exhibit both low thermal expansion coefficients (CTE) and high bulk moduli (BM). These alloys are highly sought after in various technologies where dimensional stability, mechanical strength, and thermal resistance are paramount. For instance, alloys such as NbTiVZr, known for its high yield strength and low CTE at elevated temperatures \cite{10.3389/fmats.2020.00172}, and ZrNbAl, recognized for its low CTE and robust mechanical properties \cite{10.1088/0256-307x/35/8/086501}, are ideal for engineering applications that demand thermal dimensional stability and strength. Furthermore, the fine-tuning of CTE is crucial when considering self-compatible base alloys and thermal/environmental protection systems in turbine blade applications \cite{karaoglanli2017state}. The extensive design space of high-entropy alloys (HEAs) offers a rich landscape for discovering new alloys with these combined properties; however, the complexity of this space necessitates intelligent and informed experimentation.

This study aims to demonstrate the superiority of Deep Gaussian Process Bayesian Optimization (DGP-BO) and Multi-task Gaussian Process Bayesian Optimization (MTGP-BO) over conventional Gaussian Process Bayesian Optimization (cGP-BO) within the FeCrNiCoCu HEA system, focusing on two specific sets of objectives. Specifically, the proposed framework will be used to explore compositions with both low CTE-high BM and high CTE-high BM in the vast HEA design space using high-throughput atomistic simulations. 

Our ultimate objective is to demonstrate that advanced kernel-based Bayesian Optimization (BO) methods, such as Deep Gaussian Process BO (DGP-BO) and Multi-Task Gaussian Process BO (MTGP-BO), can effectively leverage and learn from correlated properties, thereby outperforming conventional Gaussian Process BO (cGP-BO), which lacks the ability to utilize shared information across correlated properties. Given that material properties are inherently correlated and share information among themselves, we hypothesize that methods like DGP-BO and MTGP-BO, which explicitly exploit mutual information across different properties (or 'tasks' in the context of machine learning), will exhibit markedly superior performance compared to methods that are agnostic to these correlations. Since all realistic materials discovery problems are inherently multi-objective and properties are interrelated, the methods demonstrated in this work have broad applicability in a diverse set of problems associated with materials discovery and optimization.

\section{Methods}
\subsection{Mathematical Basis of GP and MTGP}

In materials discovery applications, GP and MTGP models offer powerful frameworks for predictive modeling and optimization. GPs are a robust tool for modeling objective functions in multi-objective BO, providing a probabilistic approach to regression that quantifies uncertainty in predictions \cite{rasmussen2006gaussian}. Traditional BO algorithms often model each objective function independently using separate GPs, thus assuming no correlation between them. This independent modeling approach can overlook potential interrelationships among objective functions. In contrast, the MTGP approach models all objective functions jointly, capturing any correlations and leveraging shared information to improve predictive accuracy \cite{bonilla2008multi}.

Consider a GP constructed for an objective function \( f(x) \). The prediction at an unobserved location \( \mathbf{x} \) given \( N \) previously observed data points denoted by \( \{ \mathbf{X}_N, \mathbf{y}_N \} \), where \( \mathbf{X}_N = (x_1, x_2, \ldots, x_N) \) and \( \mathbf{y}_N = (f(x_1), f(x_2), \ldots, f(x_N)) \), is given as:

\[
f_{\text{GP}}(x) \mid \mathbf{X}_N, \mathbf{y}_N \sim \mathcal{N}\left( \mu(x), \sigma^2(x) \right)
\]

where the mean \( \mu(x) \) and variance \( \sigma^2(x) \) are defined as:

\[
\mu(x) = \mathbf{K}(\mathbf{X}_N, x)^\top \left[ \mathbf{K}(\mathbf{X}_N, \mathbf{X}_N) + \sigma^2 \mathbf{I} \right]^{-1} \mathbf{y}_N
\]

\[
\sigma^2(x) = k(x, x) - \mathbf{K}(\mathbf{X}_N, x)^\top \left[ \mathbf{K}(\mathbf{X}_N, \mathbf{X}_N) + \sigma^2 \mathbf{I} \right]^{-1} \mathbf{K}(\mathbf{X}_N, x)
\]

Here, \( k \) is the kernel function, \( \mathbf{K}(\mathbf{X}_N, \mathbf{X}_N) \) is an \( N \times N \) matrix where the \( (m,n) \)-th entry is \( k(x_m, x_n) \), and \( \mathbf{K}(\mathbf{X}_N, x) \) is an \( N \times 1 \) vector with the \( m \)-th entry being \( k(x_m, x) \). The term \( \sigma^2 \) represents experimental noise. The kernel function captures the correlation between observations based on their relative distances \cite{duvenaud2014automatic}---these distances may not necessarily be defined in a Euclidean space. A common choice for the kernel function is the squared exponential kernel:

\[
k_x(x, x') = \exp\left( -\sum_{h=1}^d \frac{(x_h - x'_h)^2}{2l_h^2} \right)
\]

where \( d \) is the dimensionality of the input space, and \( l_h \) is the characteristic length-scale determining the strength of correlation in the \( h \)-th dimension.

The MTGP model extends this by jointly modeling multiple objective functions, taking into account their inter-task correlations \cite{alvarez2012kernels}. To achieve this, we define a task-specific covariance matrix \( \mathbf{K}_f \) to capture the similarities between tasks. The overall covariance function for the MTGP model is constructed using the Kronecker product of the task covariance matrix \( \mathbf{K}_f \) and the input covariance matrix \( \mathbf{K}_x \):

\[
\mathbf{K} = \mathbf{K}_f \otimes \mathbf{K}_x, \quad k = \mathbf{K}_f \otimes k_x
\]

Here, \( \otimes \) denotes the Kronecker product. The variable \( \mathbf{y}_N \) in the equations is a column vector, where each task's data is ordered in columns and stacked vertically.

This MTGP model allows us to capture the correlations between different objective functions, thereby improving the accuracy and efficiency of the optimization framework. By leveraging the shared information between tasks, the MTGP can enhance the performance of the design optimization process, leading to more accurate predictions and faster convergence to optimal solutions \cite{swersky2013multi}. Inference in the MTGP model follows the standard GP equations for the mean and variance of the predictive distribution, extended to the multi-task setting. The mean prediction for a new data point \( x^* \) for task \( l \) is given by:

\[
f_l(x^*) = (\mathbf{k}_f^l \otimes \mathbf{k}_x^*)^\top \Sigma^{-1} \mathbf{y}
\]

where \( \mathbf{k}_f^l \) selects the \( l \)-th column of \( \mathbf{K}_f \), \( \mathbf{k}_x^* \) is the vector of covariances between the test point \( x^* \) and the training points, and \( \Sigma \) is the combined covariance matrix given by \( \Sigma = \mathbf{K}_f \otimes \mathbf{K}_x + \mathbf{D} \otimes \mathbf{I} \), where \( \mathbf{D} \) is a diagonal matrix representing noise variances.
Here, \( y \) represents the observed data across all tasks, stacked appropriately. The use of MTGP allows for the transfer of information between tasks, leveraging the inter-task correlations to improve predictive performance, especially when data is sparse for some tasks. This joint modeling approach is particularly beneficial in materials discovery, where the properties of materials are often correlated, and exploiting these correlations can lead to more efficient and accurate predictions \cite{bonilla2007multi,khatamsaz2023multi}.

The hyperparameters of the MTGP model, including the parameters of the kernel functions and the task covariance matrix, are learned by maximizing the marginal likelihood of the observed data \cite{rasmussen2006gaussian}. This can be done using gradient-based optimization methods, ensuring the positive semi-definiteness of \( \mathbf{K}_f \) through parametrization such as the Cholesky decomposition \cite{alvarez2012kernels}. Overall, the MTGP model provides a flexible and powerful approach for multi-objective optimization, capturing inter-task correlations and enhancing the efficiency of the design framework.

\subsection{Mathematical Basis of DGP}

Deep Gaussian Processes (DGPs) extend the concept of GP to hierarchical models, providing a more powerful and flexible approach to modeling complex data \cite{damianou2013deep}. A DGP consists of multiple layers of latent variables, where each layer is modeled by a GP. This hierarchical structure allows for the representation of more abstract features at higher layers, which can improve predictive performance in tasks with complex underlying structures \cite{salimbeni2017doubly}.

In a standard GP, we model a set of training input-output pairs, \(\{\mathbf{X}, \mathbf{Y}\}\), using a latent function \( f(x) \) drawn from a GP prior, and we infer the distribution of \( f \) given the data. Observed data points are denoted by  \(\{\mathbf{X}, \mathbf{Y}\}\), where \( \mathbf{X} = (x_1, x_2, \ldots, x_N) \) and \( \mathbf{Y} = (y_1, y_2, \ldots, y_N) \). In the case of the DGP, the inputs to the GP at each layer are governed by the outputs of the GP from the previous layer. Formally, if we consider a DGP with \( H \) hidden layers, the generative process can be described as:

\[
\begin{aligned}
    &\mathbf{y}_n = f^Y(\mathbf{h}_n^1) + \epsilon^Y_n, & \epsilon^Y_n \sim \mathcal{N}(0, \sigma^2_Y I), \\
    &\mathbf{h}_n^1 = f^1(\mathbf{h}_n^2) + \epsilon^1_n, & \epsilon^1_n \sim \mathcal{N}(0, \sigma^2_1 I), \\
    &\vdots \\
    &\mathbf{h}_n^{H-1} = f^{H-1}(\mathbf{h}_n^H) + \epsilon^{H-1}_n, & \epsilon^{H-1}_n \sim \mathcal{N}(0, \sigma^2_{H-1} I), \\
    &\mathbf{h}_n^H = \mathbf{x}_n,
\end{aligned}
\]
where \( \mathbf{y}_n \) are the observed data points, \( \mathbf{h}_n^i \) are the latent variables at the \( i \)-th layer, and \( \mathbf{x}_n \) are the discrete data points used as the inputs for the highest layer in the hierarchy. Each function \( f^i \) is modeled as a GP with an appropriate covariance function \cite{bui2016deep}.

Training a DGP involves marginalizing over the latent variables in each layer, which is analytically intractable. To address this, we use a variational approach to approximate the marginal likelihood \cite{titsias2009variational}. The variational distribution \( Q \) is introduced to approximate the true posterior distribution of the latent variables. The variational lower bound on the marginal likelihood is given by:

\[
\mathcal{L} = \int Q \log \frac{p(\mathbf{Y}, \{\mathbf{H}^i\}, \{\mathbf{F}^i\})}{Q},
\]
where \( Q \) is factorized as:

\[
Q = \prod_{i=1}^H \left[ p(\mathbf{F}^i|\mathbf{U}^i, \mathbf{H}^{i+1}) q(\mathbf{U}^i) q(\mathbf{H}^i) \right],
\]
and \( \mathbf{U}^i \) are inducing variables for the \( i \)-th layer, which is a smaller set of data points designed to approximate the posterior for each layer while managing computational complexity. The terms in the variational lower bound can be computed analytically, allowing for efficient optimization of the model parameters and variational distributions \cite{hensman2013gaussian}.

This Bayesian training approach not only provides a principled way to learn the model parameters but also enables automatic determination of the appropriate structure for the deep hierarchy through the use of automatic relevance determination (ARD) priors \cite{neal2012bayesian}. The ARD priors help in identifying the most relevant dimensions at each layer, facilitating effective model complexity control.

Overall, DGPs offer a robust framework for capturing complex dependencies in data, making them particularly suitable for applications in materials discovery where the relationships between variables can be highly nonlinear and intricate \cite{damianou2013deep, damianou2015deep}.

\subsection{Mathematical Basis of Expected Hypervolume Improvement (EHVI) and Upper Confidence Bound (UCB)}

In BO, acquisition functions guide the search for the optimal solution by balancing exploration and exploitation. Two popular acquisition functions are the Upper Confidence Bound (UCB) and Expected Hypervolume Improvement (EHVI).

\subsubsection{Upper Confidence Bound (UCB)}

The UCB acquisition function selects the next query point by maximizing a confidence interval around the prediction. It is defined as:
\[
\text{UCB}(x) = \mu(x) + \beta \sigma(x)
\]
where \(\mu(x)\) is the mean prediction at point \(x\), \(\sigma(x)\) is the standard deviation, and \(\beta\) is a parameter that controls the trade-off between exploration (higher \(\beta\)) and exploitation (lower \(\beta\)) \cite{wilson2018maximizing}. By considering both the predicted mean and the uncertainty, UCB ensures that regions with high uncertainty are explored, which helps in finding the global optimum efficiently \cite{srinivas2009gaussian}.

\subsubsection{Expected Hypervolume Improvement (EHVI)}

In multi-objective BO, EHVI is a crucial metric for evaluating the potential of new candidate solutions in expanding the Pareto front of an optimization problem. Here, we present the mathematical formulation of EHVI, based on the methodologies discussed in the literature \cite{emmerich2008computation, yang2019expected}.

\subsubsection{Definition of Dominance and Hypervolume}

Consider a multi-objective optimization problem with \(m\) objectives:

\begin{equation}
f_1(x) \to \min, \ldots, f_m(x) \to \min, \quad x \in X
\end{equation}
Let \( \mathbf{y} \in \mathbb{R}^m \) be a vector of objective values. A point \( \mathbf{y} \) is said to dominate another point \( \mathbf{y'} \in \mathbb{R}^m \), denoted as \( \mathbf{y} \prec \mathbf{y'} \), if:
\begin{equation}
\forall i \in \{1, \ldots, m\} : y_i \leq y'_i \quad \text{and} \quad \mathbf{y} \neq \mathbf{y'}
\end{equation}
The hypervolume \( S(P, \mathbf{y}_{\text{ref}}) \) of a set \( P \) with respect to a reference point \( \mathbf{y}_{\text{ref}} \) is defined as:
\begin{equation}
S(P, \mathbf{y}_{\text{ref}}) = \int_{\mathbf{y} \in \mathbb{R}^m} T(P \prec \mathbf{y} \prec \mathbf{y}_{\text{ref}}) \, d\mathbf{y}
\end{equation}
where \( T \) is an indicator function that returns 1 if \( P \prec \mathbf{y} \prec \mathbf{y}_{\text{ref}} \) and 0 otherwise \cite{zitzler2003performance}.

\subsubsection{Improvement in Hypervolume}

Given a new candidate solution \( \mathbf{x} \) with predicted objective values following an independent multivariate Gaussian distribution with mean vector \( \boldsymbol{\mu} \) and standard deviation vector \( \boldsymbol{\sigma} \), the improvement in hypervolume is measured by the increase in the hypervolume of the Pareto front when \( \mathbf{x} \) is added to the population \( P \):

\begin{equation}
I(\mathbf{y}, P) = S(P \cup \{\mathbf{y}\}) - S(P)
\end{equation}

\subsubsection{Computation of EHVI}

The EHVI integrates the improvement in hypervolume over the distribution of the predicted values:
\begin{equation}
\text{EHVI}(\mathbf{x}) = \int_{\mathbf{y} \in \mathbb{R}^m} I(\mathbf{y}, P) \cdot \text{PDF}_{\mathbf{x}}(\mathbf{y}) \, d\mathbf{y}
\end{equation}
where \( \text{PDF}_{\mathbf{x}} \) is the probability density function of the predicted values of \( \mathbf{x} \) \cite{emmerich2011hypervolume}.
To compute the EHVI, the integration region \( \mathbb{R}^m \) is partitioned into a set of interval boxes using a grid defined by the current Pareto front \( P \) and the reference point \( \mathbf{y}_{\text{ref}} \). Each grid cell \( C(i_1, \ldots, i_m) \) has lower and upper bounds \( \mathbf{l}(i_1, \ldots, i_m) \) and \( \mathbf{u}(i_1, \ldots, i_m) \), respectively. The contribution of each active grid cell to the EHVI is computed by:
\begin{equation}
\delta(i_1, \ldots, i_m) = \left( \prod_{j=1}^m \delta_j(i_1, \ldots, i_m) \right) - S^+ \prod_{i=1}^m \left( \Phi\left( \frac{u_i - \mu_i}{\sigma_i} \right) - \Phi\left( \frac{l_i - \mu_i}{\sigma_i} \right) \right)
\end{equation}
where:
\begin{equation}
\delta_j(i_1, \ldots, i_m) = \Psi(v_j(i_1, \ldots, i_m), u_j(i_1, \ldots, i_m), \mu_j, \sigma_j) - \Psi(v_j(i_1, \ldots, i_m), l_j(i_1, \ldots, i_m), \mu_j, \sigma_j)
\end{equation}
and \( \Psi(a, b, \mu, \sigma) \) is given by:
\begin{equation}
\Psi(a, b, \mu, \sigma) = \sigma \phi\left( \frac{b - \mu}{\sigma} \right) + (a - \mu) \Phi\left( \frac{b - \mu}{\sigma} \right)
\end{equation}
Here, \( \phi \) and \( \Phi \) denote the probability density function and the cumulative distribution function of the standard normal distribution, respectively \cite{yang2019expected}.
Finally, the EHVI is obtained by summing the contributions of all active grid cells:
\begin{equation}
\text{EHVI}(\mathbf{x}) = \sum_{C(i_1, \ldots, i_m) \in C^+} \delta(i_1, \ldots, i_m)
\end{equation}

This formulation allows for an efficient and accurate computation of EHVI, enabling its use in multi-objective Bayesian optimization for alloy design problems \cite{emmerich2016multicriteria}.

Both UCB and EHVI provide robust frameworks for decision-making in Bayesian optimization, with UCB being straightforward and suitable for single-objective problems, and EHVI being powerful for multi-objective optimization by effectively balancing the exploration of the Pareto front \cite{hernandez2016predictive}.

\subsection{ Isotopic vs. Heterotopic Data in DGP-BO}
\begin{figure}[H]
\centering
\includegraphics[width=\linewidth]{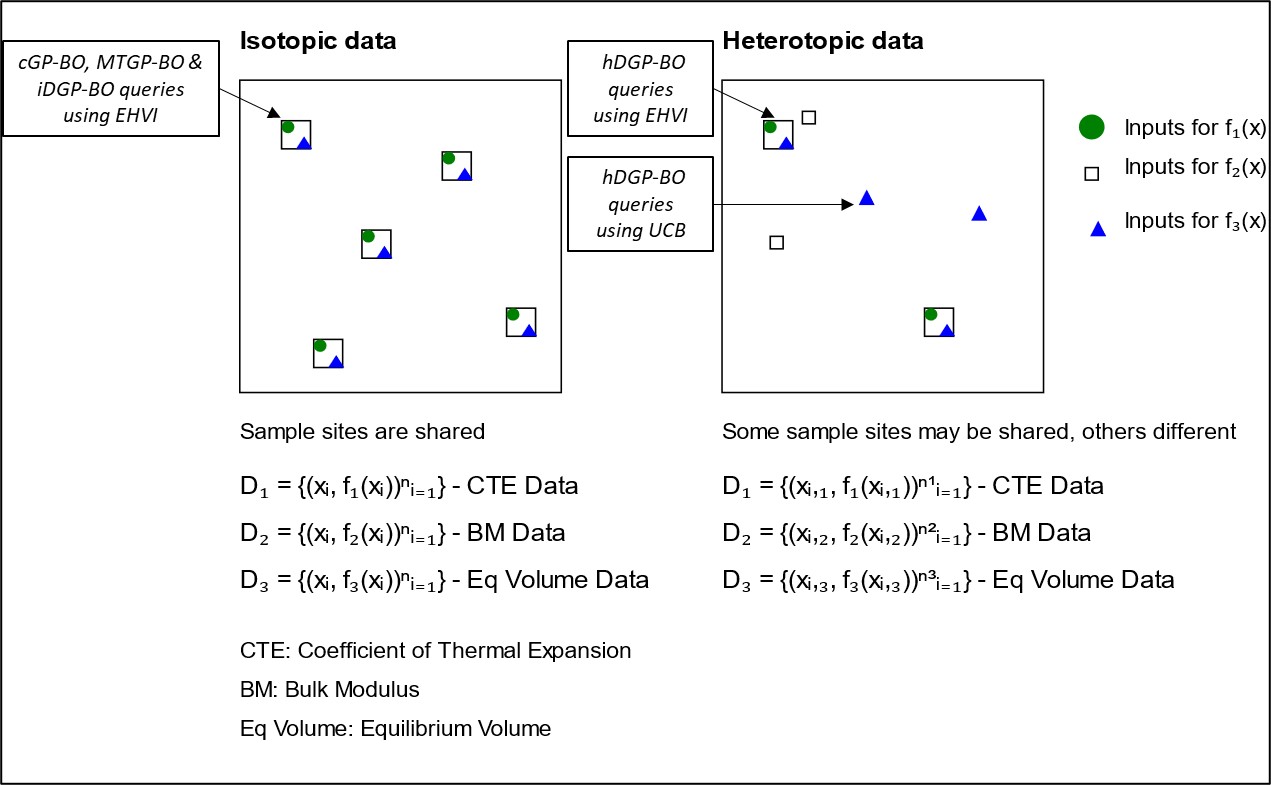}
\caption{Comparison between Isotopic and Heterotopic Data in the context of DGP. In Isotopic data, sample sites are shared across tasks, while in Heterotopic data, sample sites may differ between tasks. This difference has significant implications for the performance of Bayesian Optimization models, particularly in iDGP-BO and hDGP-BO.}
\label{fig:isotopic_heterotopic}
\end{figure}

In BO with DGPs, the distinction between isotopic and heterotopic sample sites(query sites or data points) is crucial for understanding the operational differences between Isotopic DGP-BO (iDGP-BO) and Heterotopic DGP-BO (hDGP-BO)\cite{alvarez2012kernels}.

\textbf{Isotopic data} refers to scenarios where sample sites are shared across all tasks. This means that the same input locations are used for every task, simplifying the modeling process since the correlation structure between tasks can be more easily captured due to the shared sampling sites. This is the approach taken by iDGP-BO in this work, which assumes that the tasks are closely related and can be effectively modeled using shared sample points or query sites. While this assumption reduces complexity and improves computational efficiency, it may miss task-specific nuances that arise from differing sample sites, potentially leading to suboptimal performance in complex, real-world scenarios where such assumptions do not hold. Mathematically it can be described as,
\[
\mathcal{D}_1 = \{(x_i, f_1(x_i))\}_{i=1}^N, \quad \mathcal{D}_2 = \{(x_i, f_2(x_i))\}_{i=1}^N
\]
where \( x_i \) represents the shared sample locations, and \( f_1(x_i) \) and \( f_2(x_i) \) represent the outputs for the respective tasks at those locations.

\textbf{Heterotopic data}, on the other hand, allows query sites to differ between tasks, providing greater flexibility in modeling. This is the foundation of hDGP-BO , where each task has its own set of sampling sites. This approach leverages the additional flexibility provided by different query locations to better capture task-specific correlations, making it possible to exploit these for more effective optimization. By allowing for different sample sites, hDGP-BO can perform more nuanced and accurate predictions, particularly in cases where the tasks exhibit significant variability in their data distributions. This added complexity requires more sophisticated modeling but can result in superior performance, as evidenced in our experiments. It can be expressed as,
\[
\mathcal{D}_1 = \{(x_{i,1}, f_1(x_{i,1}))\}_{i=1}^{N_1}, \quad \mathcal{D}_2 = \{(x_{i,2}, f_2(x_{i,2}))\}_{i=1}^{N_2}
\]
where \( x_{i,1} \) and \( x_{i,2} \) represent different sample locations for each task.

\textbf{iDGP-BO} uses a DGP that assumes all tasks share the same sampling sites. Here, all the queries across all the task are always made on same query locations based on EHVI acquisition function. This isotopic approach simplifies the correlation structure between tasks, making it easier to model but potentially less flexible in handling task-specific variations. The advantage of iDGP-BO lies in its simplicity and computational efficiency, which can be beneficial in scenarios where tasks are expected to be closely related.

\textbf{hDGP-BO}, however, uses a more complex framework that allows each task to have its own sampling sites (heterotopic approach). This model leverages DGP to capture more intricate correlations between tasks, which can be particularly beneficial when tasks are less related or exhibit significant individual characteristics. hDGP-BO also incorporates an acquisition strategy where less expensive tasks can be queried more frequently, using techniques like UCB to reduce uncertainty, thereby informing the sampling strategy for more expensive tasks. This makes hDGP-BO not only more flexible but also potentially more accurate in scenarios where task-specific data is critical. In our strategy, hDGP-BO makes queries using same input points one-third of the times using EHVI. During other steps, it queries only BM and volume at different input locations, based on the highest UCB of respective tasks. In this way hDGP-BO can explore or be informed about CTE at a certain location by only querying the BM or volume at that specific location. This occurs because of the correlated structure of DGP. Thus, knowing one task at a specific input location informs the GP about associated tasks at that location without explicitly querying
those associated tasks.

In the workflow of this current work, hDGP-BO queries CTE, BM and volume 500 times using EHVI during the whole BO procedure (isotopic queries). In addition to that, hDGP-BO queries only BM and volume twice using UCB before each isotopic queries. So, in total hDGP-BO acquires CTE data 500 times and BM, volume data 1500 times. Whereas, models executing only isotopic queries (cGP-BO, MTGP-BO \& iDGP-BO) acquires CTE, BM and volume data 500 times each. hDGP-BO leverages this additional relatively inexpensive BM and volume data to carry out the optimization task in a more efficient and robust manner. 

\subsection{High-throughput Bayesian Optimization Framework}

As the underlying simulation platform for our BO, we employed the pyiron workflow framework~\cite{janssen2019pyiron2} to conduct high-throughput screening simulations for sampling the FeCrNiCoCu high entropy alloy (HEA) space. pyiron provides workflows to calculated material properties independent of the simulation engine, including thermal expansion coefficeints~(CTE), equilibrium bulk modulus~(BM) and equilibrium volume~(V). We calculate those with an atomic composition resolution of 3\%, amounting to a total of roughly 60,000 configuration to sample the composition space. For benchmarking the BO schemes the target properties were calculated for all compositions {\em a priori}. Subsequently, the BO schemes queried from this consolidated database for optimization, which accelerates benchmarking the different BO schemes and facilitates systematic performance analysis. 

We emphasize that, in a real experimental design scenario, the properties would only be evaluated at the query points in an on-the-fly fashion, a mode of operation that our workflow is designed to support. The \texttt{pyiron} workflow framework facilitates the seamless integration and automation of various simulation tools and workflows, significantly enhancing the efficiency and scalability of high-throughput screenings. Conceptually, \texttt{pyiron} functions as a high-level task-based interface to the simulation engine responsible for the material property calculations. In our case, the Large-scale Atomic/Molecular Massively Parallel Simulator (LAMMPS)~\cite{plimpton1995fast,LAMMPS} is employed to efficiently evaluate material properties across a large number of configurations. Additionally, \texttt{pyiron} provides a user-friendly Python interface for setting up, executing, storing, and analyzing high-throughput molecular dynamics (MD) simulations at large computational scales on high-performance computing (HPC) resources, which is critical for the coupling with Bayesian Optimization (BO).

To efficiently sample the composition space and ensure a realistic representation of the random nature of HEAs, we utilized Special Quasirandom Structures (SQS)~\cite{zunger1990special, wei2008special} with 32 atoms, generated with the sqsgenerator package ~\cite{GEHRINGER2023108664}. SQS are designed to mimic the perfectly random atomic arrangements of alloys in finite supercells, providing more accurate simulation results. The atomic configuration were evaluted using the embedded atom method (EAM) interatomic potential developed by Deluigi et al., which was designed for simulating FeNiCrCoCu \cite{deluigi2021simulations}. That being said, as the purpose of this study is to quantify the performance of various multi-task BO approaches, the absolute accuracy of the physical predictions obtained from the interatomic potential is not critical, as the general statistical properties of the correlations within and between tasks is expected to control performance.

For our workflow, each simulation began with rescaling the simulation cell based on the concentration-averaged nearest neighbor distance of its end members. This is followed by a structure optimization including relaxation of the simulation cell volume and internal degrees of freedom, resulting in an equilibrated structure at 0K. For calculation of equilibrium BM, these equilibrated structures are strained from -15\% to +15\% with an additional equilibration of the internal degrees of freedom. The BM is then calculated as the second derivative of these energy volume curves. Starting from the same equilibrated structure at 0K the CTE is determined by varying the simulations temperature from 15K to 1500K in 5K intervals over a 6ps molecular dynamics trajectory in the NPT ensemble~\cite{nose1984unified, hoover1985canonical} with a 1fs time step. By recording the change of volume with increasing temperature, the CTE is calculated providing insights into the thermal behavior of the alloys~\cite{schapotschnikow2009predicting}. These MD simulations are carried out in $3\times3\times3$ supercells of the 32 atoms SQS cells, resulting in 864 atoms cells to ensure statistical accuracy~\cite{allen2017computer, frenkel2001understanding}.The calculation of CTE differs from that of BM. The CTE require calculations of volume(along with energy minimization and NPT ensemble trajectories) at 297 different temperatures. Whereas, for calculation of equilibrium BM, there's no dependence on temperature unlike CTE. Thus, the cost associated with BM calculations is lower than that of CTE from a computational perspective. The CTE calculations requires \textbf{29800} more force evaluations (within a NPT ensemble), compared to the cost of calculation in case of BM. 

In total 3.5 billion atomic configurations were evaluated to construct the consolidated database for the roughly 60,000 compositions with over 750,000 individual simulation tasks being recorded in the pyiron database. This highlights the need of a workflow framework like pyiron for systematically evaluating the HEA composition space.

\section{Results and Discussion}

\subsection{High-throughput Atomistics Simulation}
\begin{figure}[H]
\centering
\includegraphics[width=1.05\linewidth]{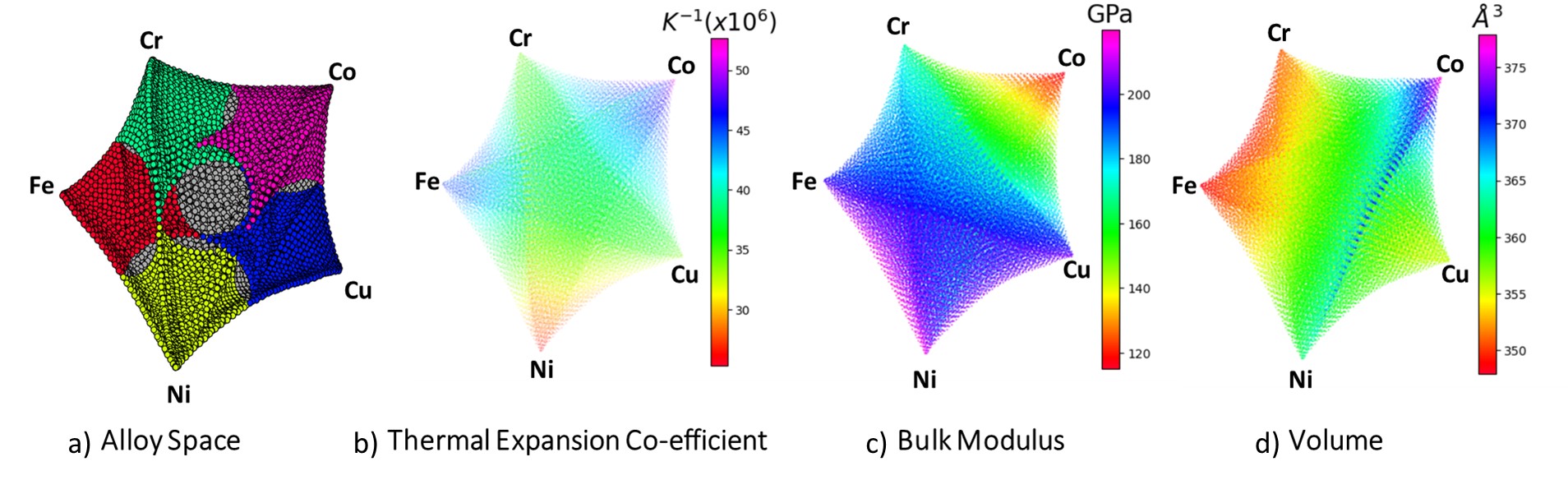}
\caption{\label{fig:maindat}The FeCrNiCoCu alloy space(a) with the thermal expansion co-efficients (b), bulk moduli (c) and equilibrium volumes (d)}
\end{figure}

The main objective of this work is to compare the performances of cGP-BO, MTGP-BO, iDGP-BO and hDGP-BO in  two specific multi-objective optimization tasks, concerning FeCrNiCoCu HEA design space. The first optimization task is to minimize CTE and maximize BM. The second optimization task involves maximization of both CTE and BM in the vast HEA space. Both of the objectives has their respective engineering applications, which will be discussed in details afterwards.

Firstly, the results of our high-throughput atomistics simulations for the FeCrNiCoCu alloy system are presented using Uniform Manifold Approximation and Projection (UMAP)  representations, as shown in \textbf{Figure \ref{fig:maindat}} \cite{mcinnes2018umap,vela2024visualizing}. UMAP, a dimension reduction technique, was chosen for its ability to preserve both local and global structure in high-dimensional data, making it particularly suited for visualizing the complex relationships in our alloy composition space.\\

UMAP offers several advantages over traditional dimension reduction methods~\cite{vela2024visualizing}:

\begin{itemize}
    \item It preserves topological structure across multiple scales, allowing for the visualization of both fine-grained local relationships and broader global patterns.
    \item It can handle non-linear relationships in the data, which are common in materials science datasets.
    \item It is computationally efficient, enabling the visualization of large datasets typical in high-throughput studies.
\end{itemize}

In our analysis, UMAP transforms the five-dimensional compositional space of the FeCrNiCoCu alloy system into a two-dimensional representation. Each point in these visualizations corresponds to a unique alloy composition, with color gradients indicating the magnitude of different material properties. This approach provides an intuitive understanding of how properties vary across the compositional landscape, allowing for rapid identification of trends, clusters, and potential high-performance regions. These plots effectively capture the complex interplay between composition and properties, offering insights that might be obscured in traditional analysis methods. The use of UMAP allows us to visually explore correlations between different properties and identify compositional regions of interest for further investigation.

\textbf{Figure \ref{fig:maindat}} reports the properties of the FeCrNiCoCu alloy space in UMAP scheme, calculated using the workflow described in the last section. \textbf{Figure \ref{fig:maindat}(a)} reports the mapping of the compositional space of the FeCrNiCoCu alloy system onto a UMAP. Colors indicate region where the alloy contains an element at more than 50\% concentration, while no single element reaches this threshold in grey regions. Pure materials are found at the vertices of the projection. 
\textbf{Figure \ref{fig:maindat}(b)} illustrates the distribution of thermal expansion coefficients across different alloy compositions in this same projection, showing significant variation and potential candidates with low CTE. \textbf{Figure \ref{fig:maindat}(c)} shows the bulk moduli for the same set of alloys, which is crucial for identifying compositions with high mechanical strength. Finally, \textbf{Figure \ref{fig:maindat}(d)} depicts the equilibrium volumes, providing additional context for the structural properties of the alloys. Note that this projection does not guarantee continuity of the various properties in the 2D plane. It should therefore be taken primarily as a visualization aid.

\begin{figure}[H]
\centering
\includegraphics[width=1.05\linewidth]{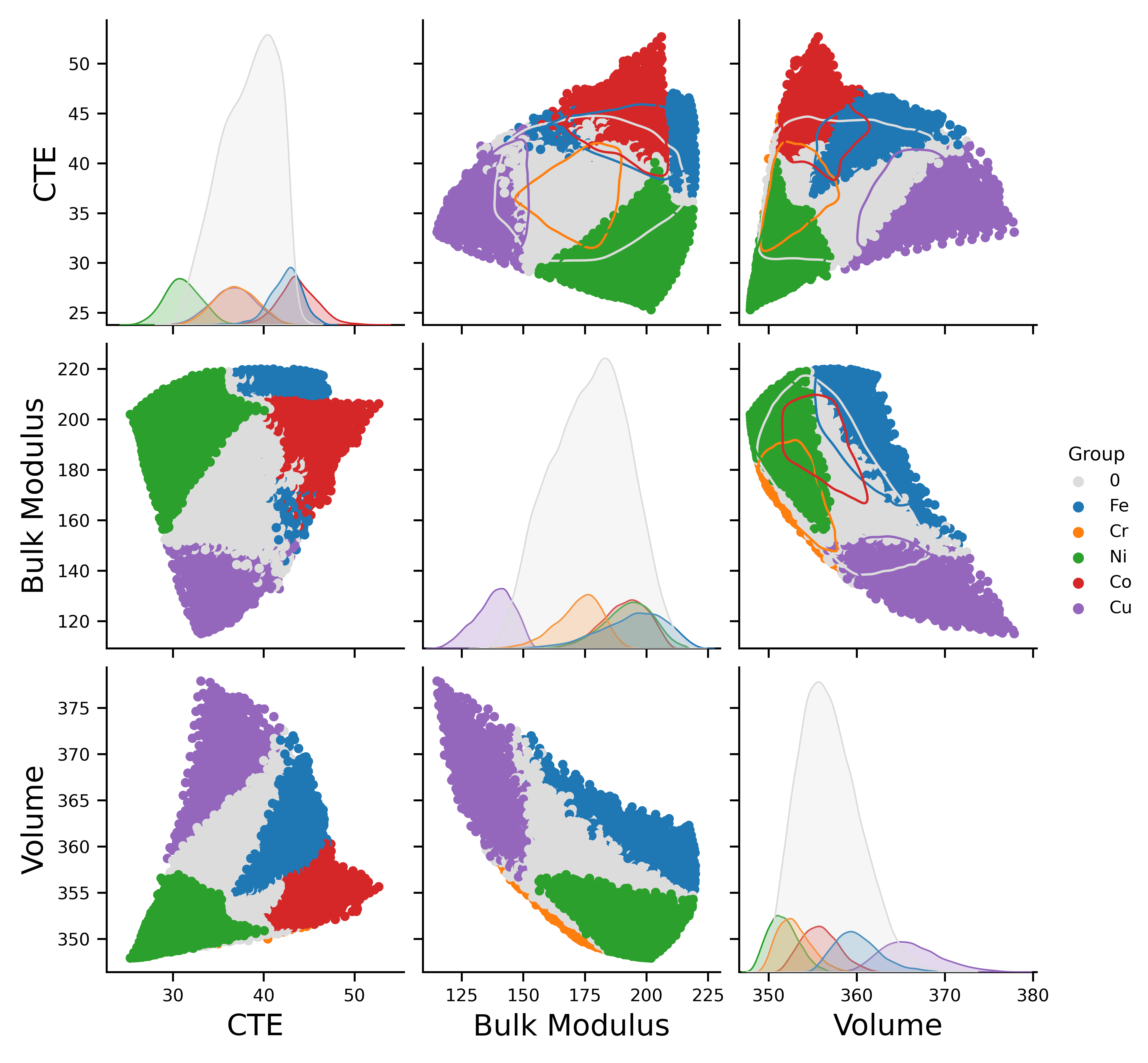}
\caption{\label{fig:1_pairwise} Pairwise distributions of the different properties, namely TEC, BM and volume in the FeCrNiCoCu alloy space. Colored regions indicate compositions with a corresponding majority element. The plots on the diagonal report distributions of the corresponding properties. }
\end{figure}

In addition to having UMAP plots for the properties individually, attempts have been made to understand the inter-relationship between the properties. 
The pairwise plots in  \textbf{Figure \ref{fig:1_pairwise}} and the correlation table in  \textbf{Table \ref{tab:correlation_matrix}} provide a deeper understanding of the inter-dependencies between thermal expansion coefficients, bulk moduli, and volumes. \textbf{Table \ref{tab:correlation_matrix}} reports a statistical correlation analysis, including Pearson correlation coefficients, Spearman coefficients, and mutual information between CTE, BM, and volume. The Pearson and Spearman coefficients provide insights into the linear and monotonic relationships between the properties, respectively, while mutual information captures both linear and non-linear dependencies \cite{cover2012elements}.

The weak correlations observed between these properties suggest the possibility of complex non-linear relationships \cite{cover2012elements, kinney2014equitability}. It is important to note that while weak linear correlations can indicate non-linear relationships, they are not definitive proof \cite{simon2014comment}. Nonetheless, the presence of weak correlations in our high-dimensional materials data motivates the exploration of more sophisticated analysis techniques \cite{ghiringhelli2015big}. Thus, the potential for using multi-task learning models to exploit these potentially complex non-linear relationships is a motivating factor for our subsequent analysis \cite{bonilla2008multi,liu2023learning,damianou2013}.

 These correlations arise from the underlying physical mechanisms governing the behavior of HEAs, such as the complex interplay between atomic size mismatch, lattice distortion, and chemistry~ \cite{miracle2017critical,zhang2014microstructures}. Empirical models to predict the thermal expansion based on the equilibrium volume, equilibrium bulk modulus and its derivative like proposed by Moruzzi et al.~\cite{Moruzzi1988} fail for these HEAs, which highlights the need for improved regression surrogates capable of learning complex relationship like MTGP or DGP for HEAs. Finally, leveraging these correlations can lead to more efficient exploration and optimization of the HEA design space.

\begin{table}[h]
\centering
\begin{tabular}{|c|ccc|ccc|ccc|}
\hline
\multirow{2}{*}{Name} & \multicolumn{3}{c|}{TEC} & \multicolumn{3}{c|}{BM} & \multicolumn{3}{c|}{Volume} \\
\cline{2-10}
 & Pearson & Spearman & MI & Pearson & Spearman & MI & Pearson & Spearman & MI \\
\hline
TEC & 1 & 1 & 1 & 0.13 & 0.15 & 0.12 & 0.32 & 0.37 & 0.20 \\
\hline
BM & 0.13 & 0.15 & 0.12 & 1 & 1 & 1 & -0.66 & -0.56 & 0.37 \\
\hline
Volume & 0.32 & 0.37 & 0.20 & -0.66 & -0.56 & 0.37 & 1 & 1 & 1 \\
\hline
\end{tabular}
\caption{Pairwise Pearson correlation coefficient, Spearman coefficient and mutual information between CTE, BM and Volume quantities of interest.}
\label{tab:correlation_matrix}
\end{table}

\subsection{Multi-objective Bayesian Optimization with Auxiliary Tasks in FeCrNiCoCu Alloy Space: Minimization of TEC - Maximization of BM}

In this section, we characterize and compare the relative efficiency of four multi-task BO approaches: cGP-BO, MTGP-BO, iDGP-BO and hDGP-BO. We apply these approaches to multi-objective material design tasks within the FeCrNiCoCu HEA space, focusing on the simultaneous minimization of CTE and maximization of BM (also, maximization of both as a case study in the next subsection), while using equilibrium volume as an auxiliary task.

This optimization scenario is particularly relevant for advanced technological applications requiring materials with exceptional dimensional stability under varying thermal conditions while maintaining high mechanical strength. Such materials are crucial in industries like aerospace, where components must withstand extreme temperature fluctuations without compromising structural integrity. For instance, in the design of next-generation aircraft engines or hypersonic vehicles, alloys with low CTE and high BM are essential for maintaining precise geometries and resisting deformation under high-stress, high-temperature environments \cite{ miracle2017critical,senkov2018development}.

Our investigation aims to demonstrate how advanced BO techniques, particularly MTGP-BO and DGP-BO, can leverage inter-task correlations and complex, non-linear relationships in the material property space to accelerate the discovery of HEAs with these desired characteristics. By comparing these methods against  cGP-BO, we seek to quantify the potential improvements in optimization efficiency and highlight the advantages of multi-task learning approaches in navigating the vast and complex design space of HEAs. Among all the different variants of BO, our novel hDGP-BO has demonstrated the most significant improvements over all other BO variants. The primary objective in this scenario was to minimize the CTE while maximizing the BM.

\begin{figure}[H]
\centering
\includegraphics[width=1.05\linewidth]{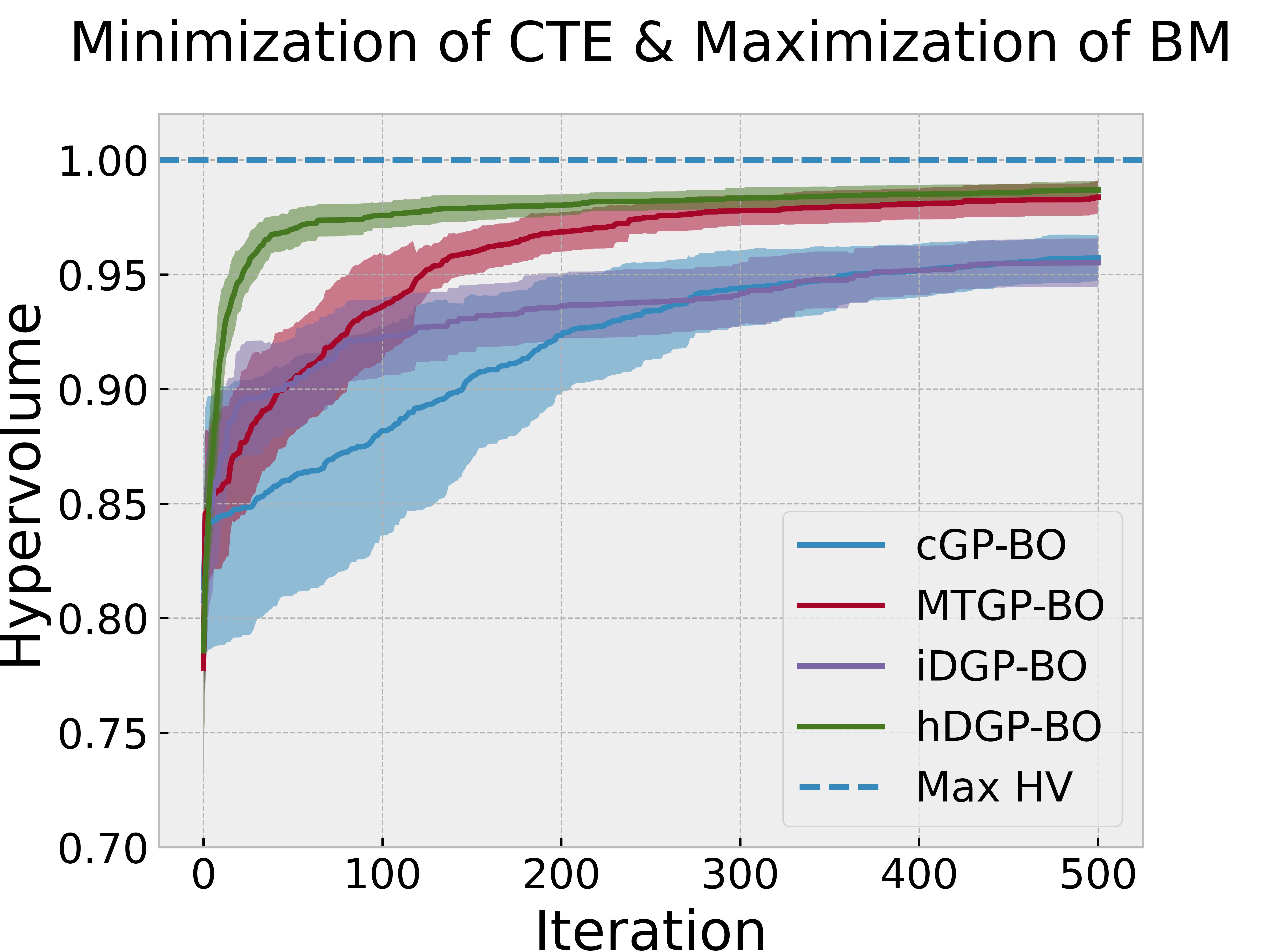}
\caption{\label{fig:hv-minmax}The hypervolume(HV) per iteration for BO based on cGP(conventional GP), MTGP(Multi-task GP), iDGP(isotopic Deep GP) and hDGP(heterotopic Deep GP)}
\end{figure}

\textbf{Figure \ref{fig:hv-minmax}} reports the evolution of the hypervolume during an iterative BO procedure based on different models, where an iteration is defined by 
a single query to all tasks using EHVI, except for the heterotopic DGP-BO (hDGP-BO)
model where one iteration of hDGP-BO comprises of two steps of queries to BM and volume using UCB, and one step of query to CTE, BM and volume using EHVI. 
So, during 500 iteration, hDGP-BO acquired 500 CTE, 1500 BM and 1500 volume data points,  whereas, cGP-BO, MTGP-BO \& iDGP-BO acquired 500 datapoints of CTE, BM and volume each. One of the objectives is to investigate whether hDGP-BO can use these extra to accelerate the mapping of the Pareto front with respect to isotopic BO variants.

\begin{figure}[H]
 \centering
 \begin{subfigure}{0.49\textwidth}
     \includegraphics[width=\textwidth]{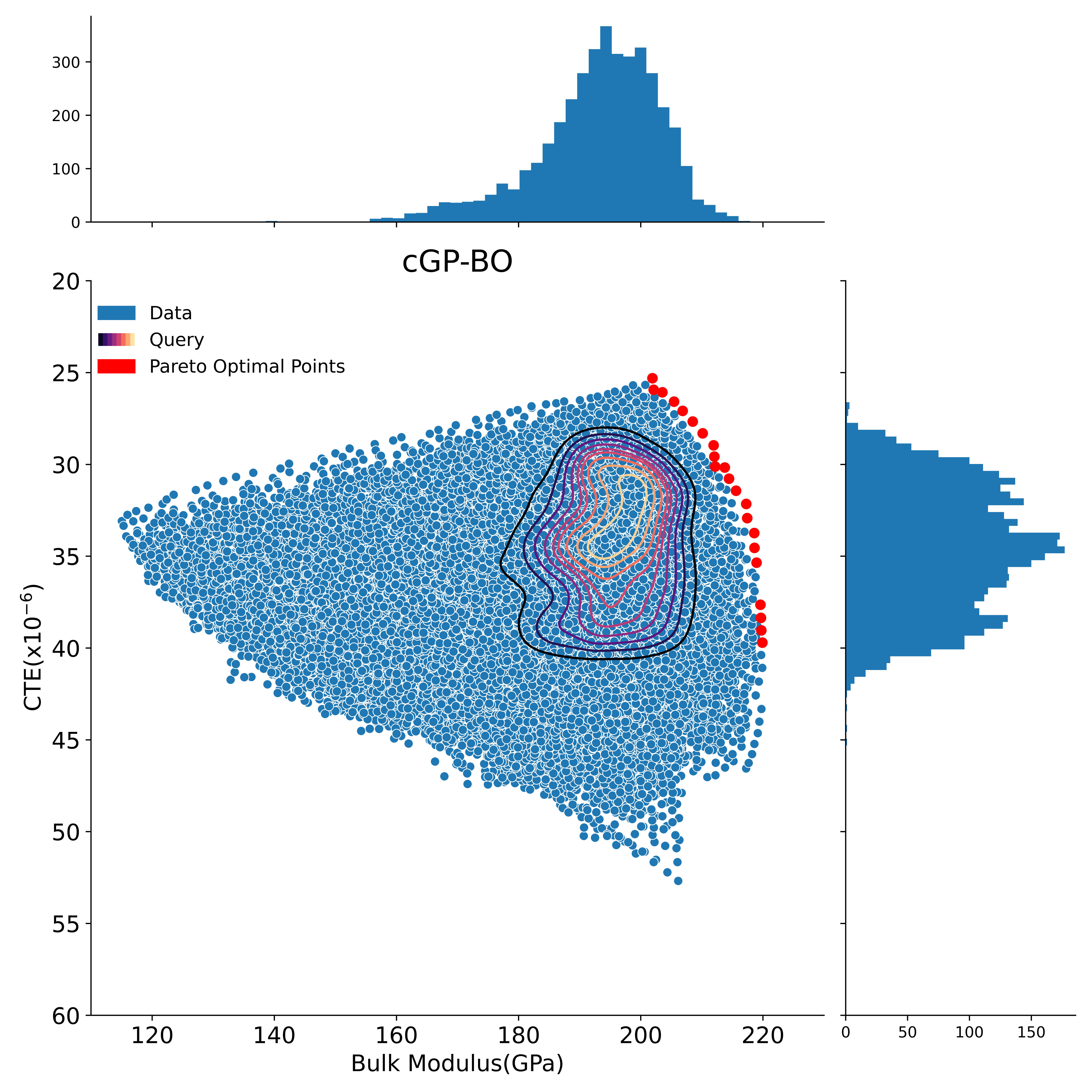}
     \caption{}
     \label{fig:queries-objective-a}
 \end{subfigure}
 \hfill
 \begin{subfigure}{0.49\textwidth}
     \includegraphics[width=\textwidth]{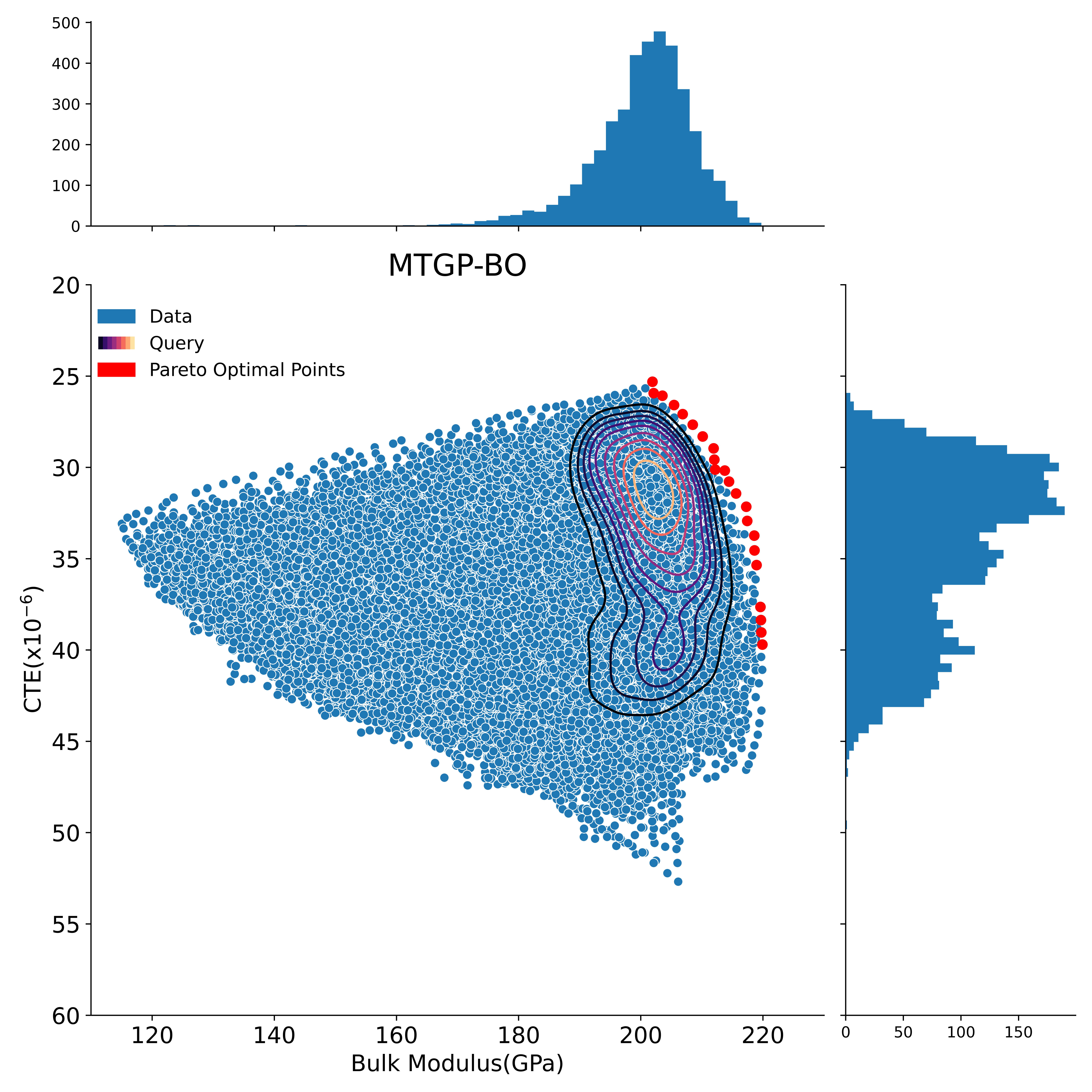}
     \caption{}
     \label{fig:queries-objective-b}
 \end{subfigure}
 
 \medskip
 \begin{subfigure}{0.49\textwidth}
     \includegraphics[width=\textwidth]{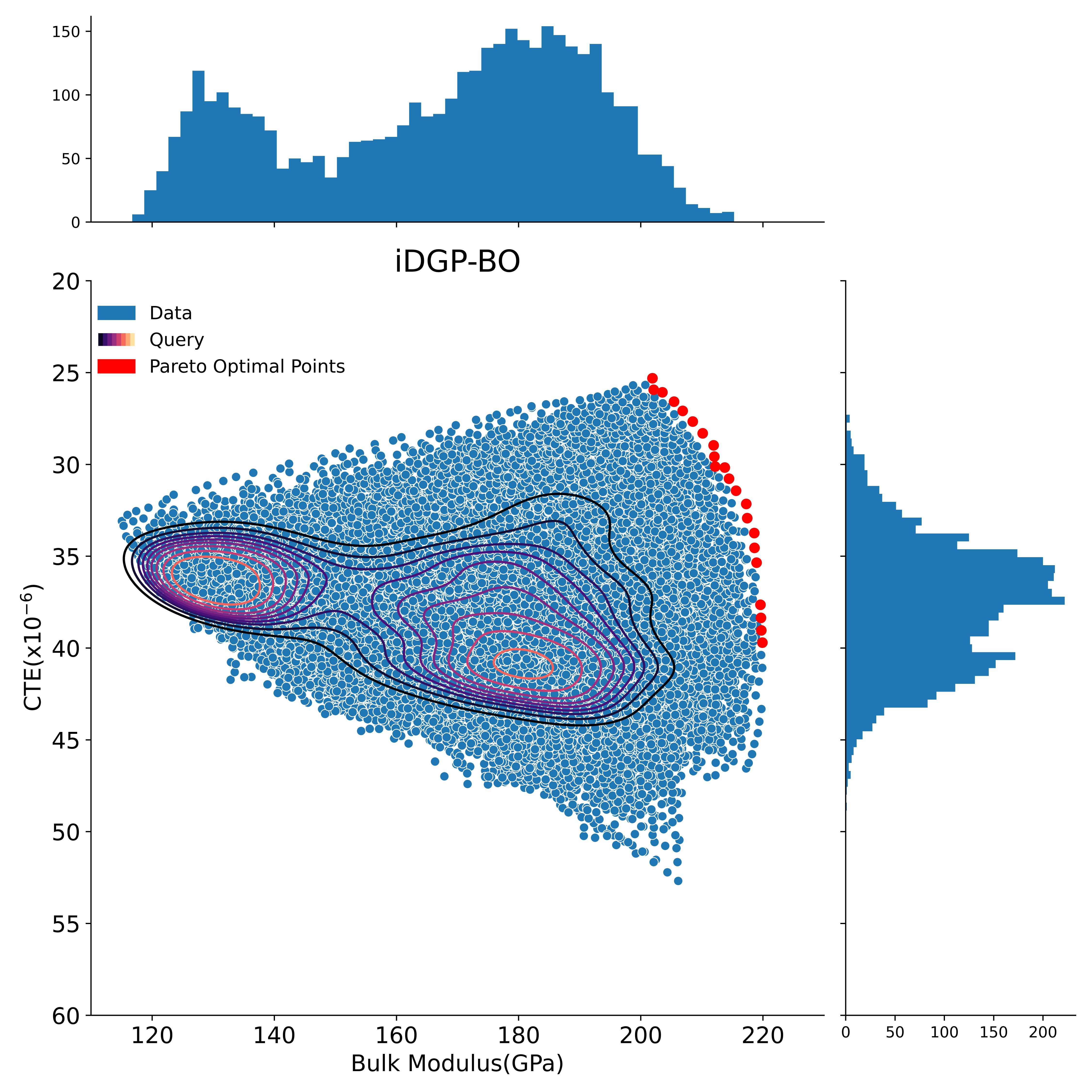}
     \caption{}
     \label{fig:queries-objective-c}
 \end{subfigure}
 \hfill
 \begin{subfigure}{0.49\textwidth}
     \includegraphics[width=\textwidth]{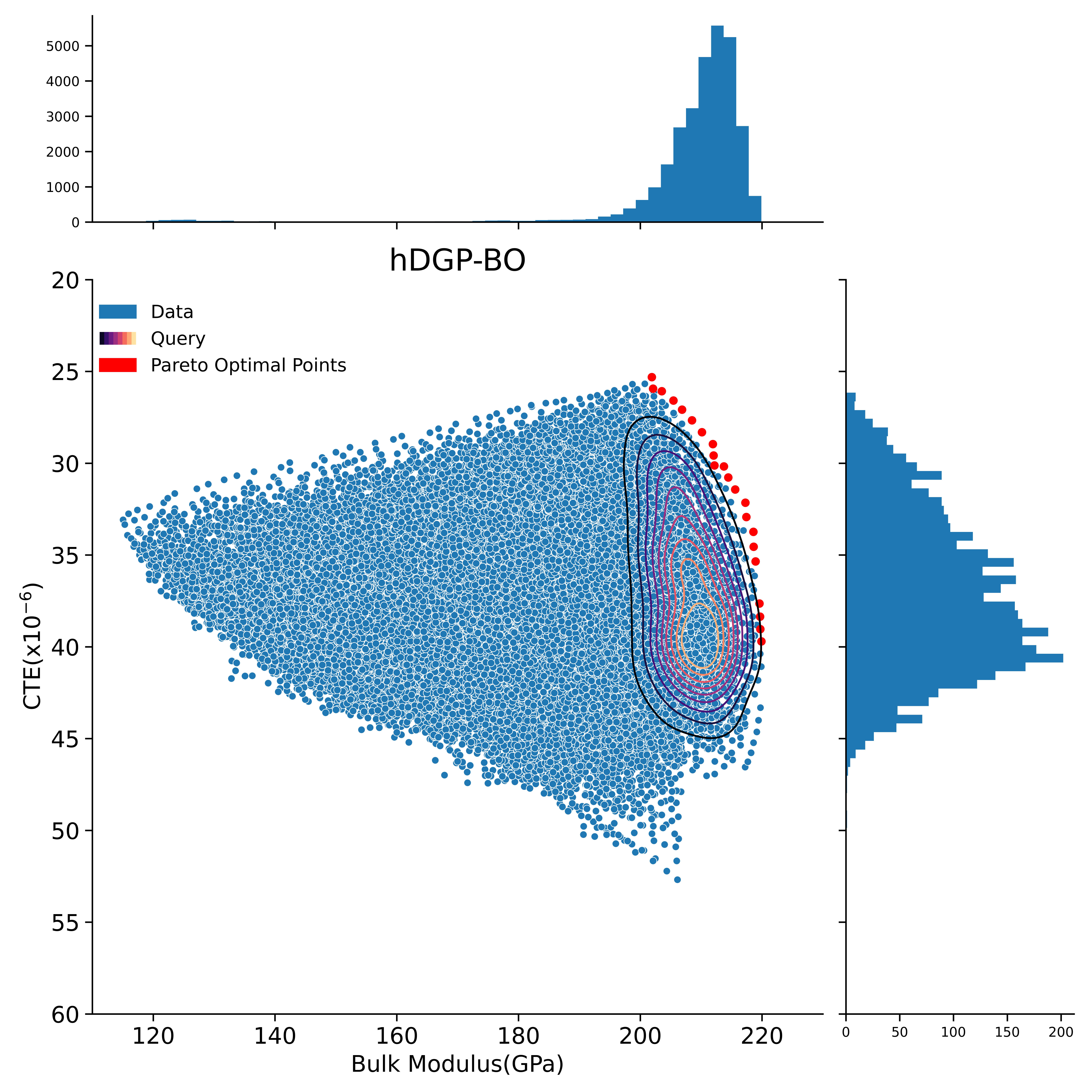}
     \caption{}
     \label{fig:queries-objective-d}
 \end{subfigure}

 \caption{The initial 200 queries in objective space by (a) cGP-BO, (b) MTGP-BO, (c) iDGP-BO and (d) hDGP-BO. Here, CTE in Y-axis is reversed, because the goal is to minimize CTE}
 \label{fig:queries-objective}
\end{figure}

\begin{figure}[H]
 \centering
 \begin{subfigure}{0.49\textwidth}
     \includegraphics[width=\textwidth]{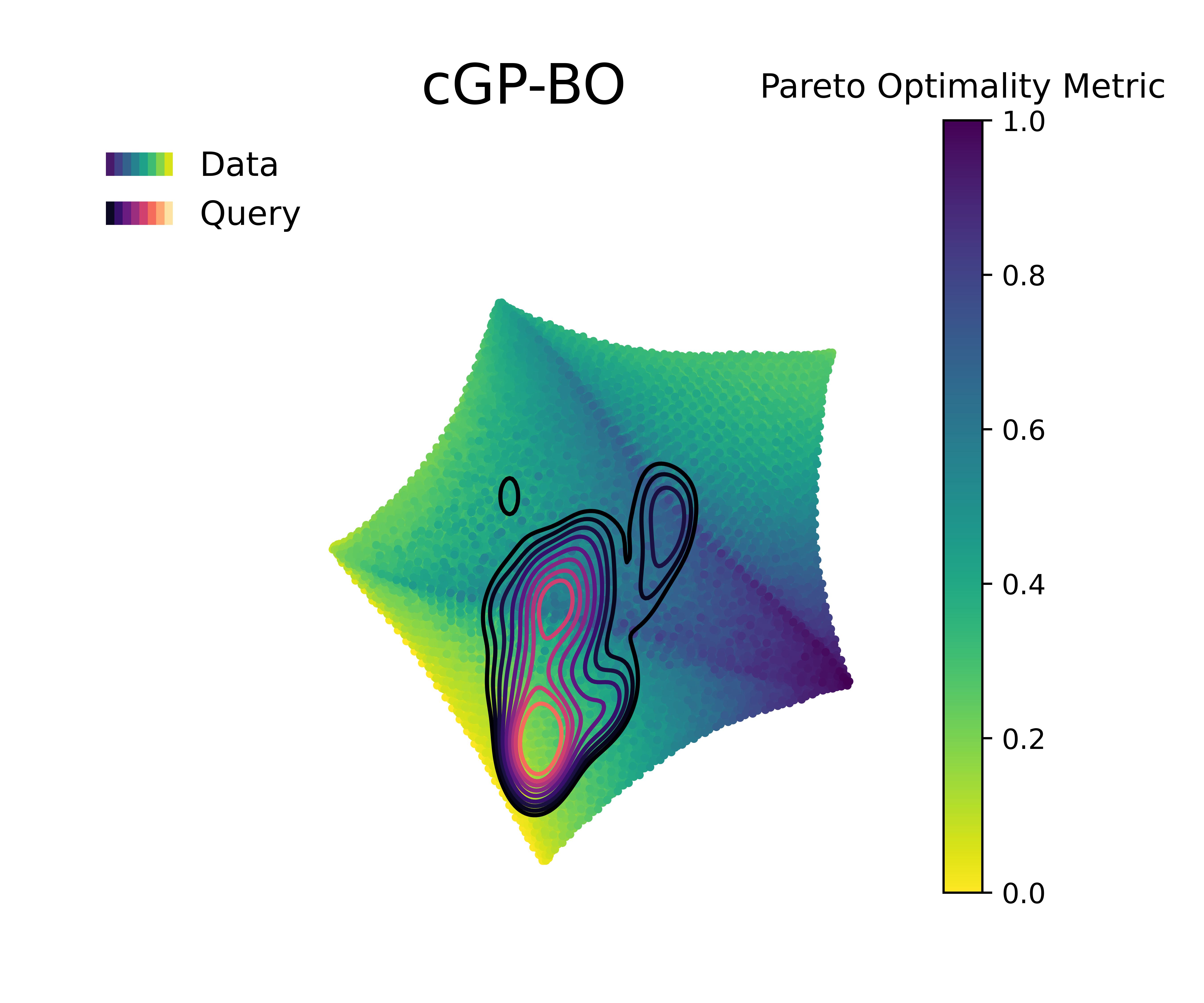}
     \caption{}
     \label{fig:queries-design-a}
 \end{subfigure}
 \hfill
 \begin{subfigure}{0.49\textwidth}
     \includegraphics[width=\textwidth]{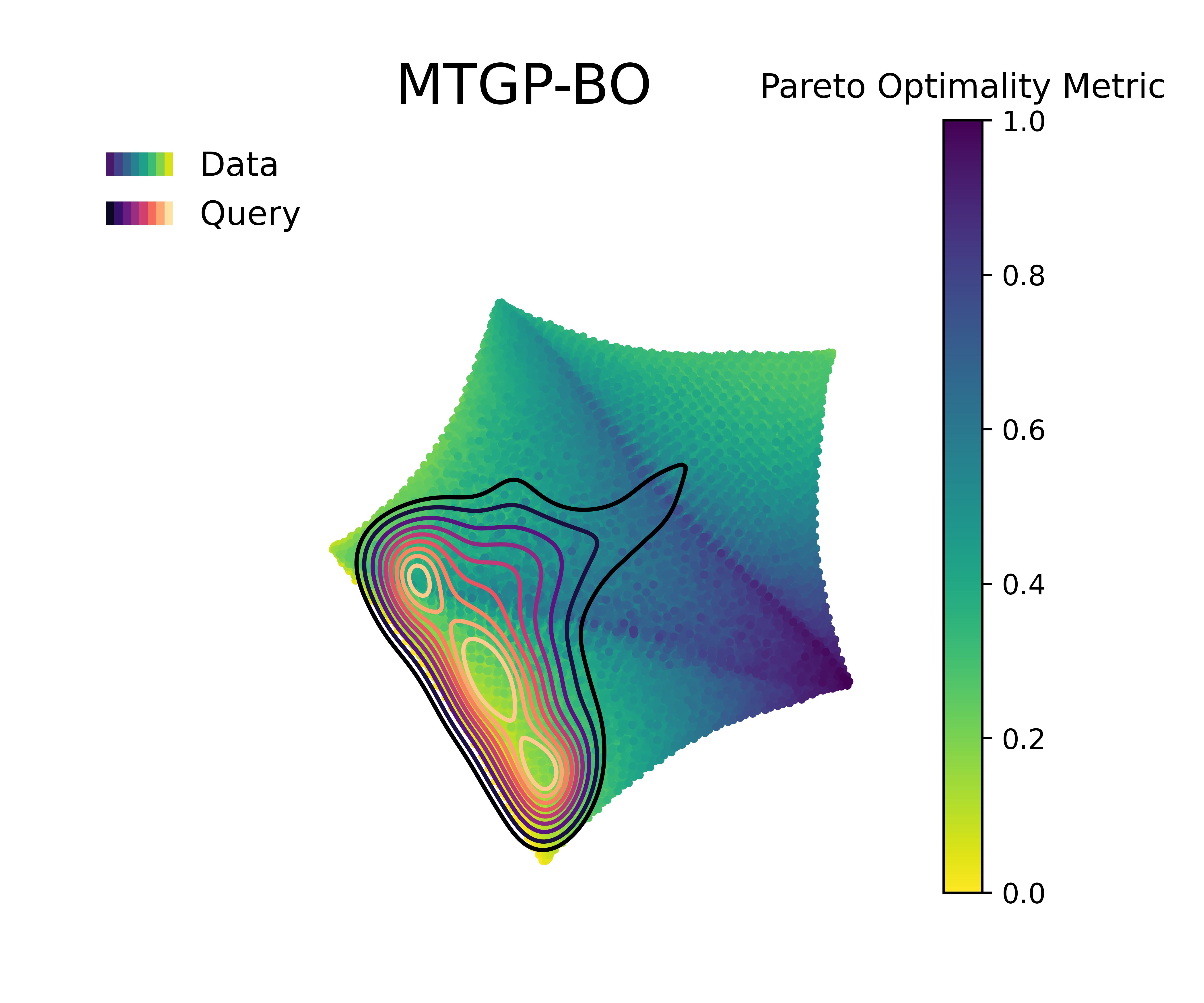}
     \caption{}
     \label{fig:queries-design-b}
 \end{subfigure}
 
 \medskip
 \begin{subfigure}{0.49\textwidth}
     \includegraphics[width=\textwidth]{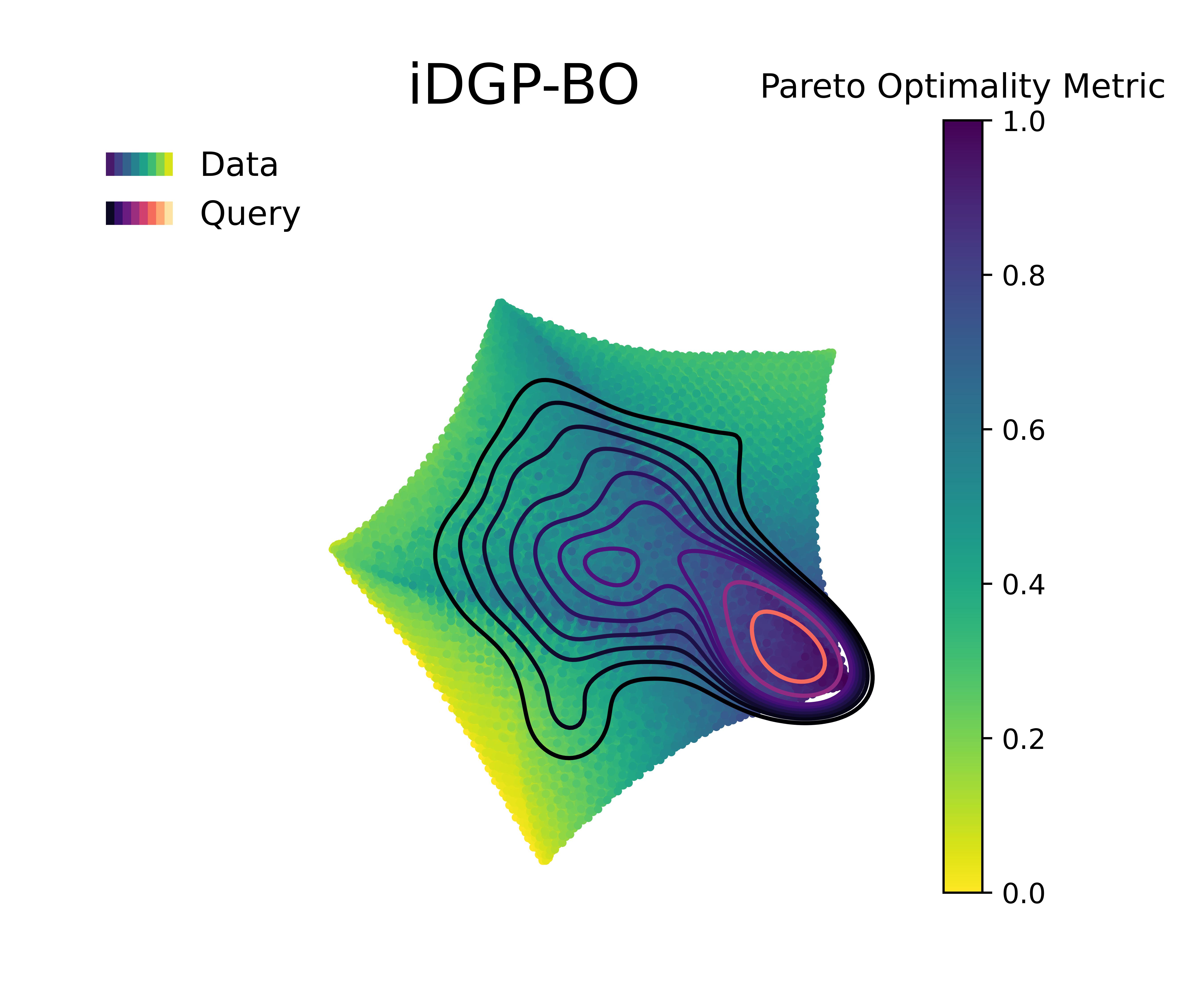}
     \caption{}
     \label{fig:queries-design-c}
 \end{subfigure}
 \hfill
 \begin{subfigure}{0.49\textwidth}
     \includegraphics[width=\textwidth]{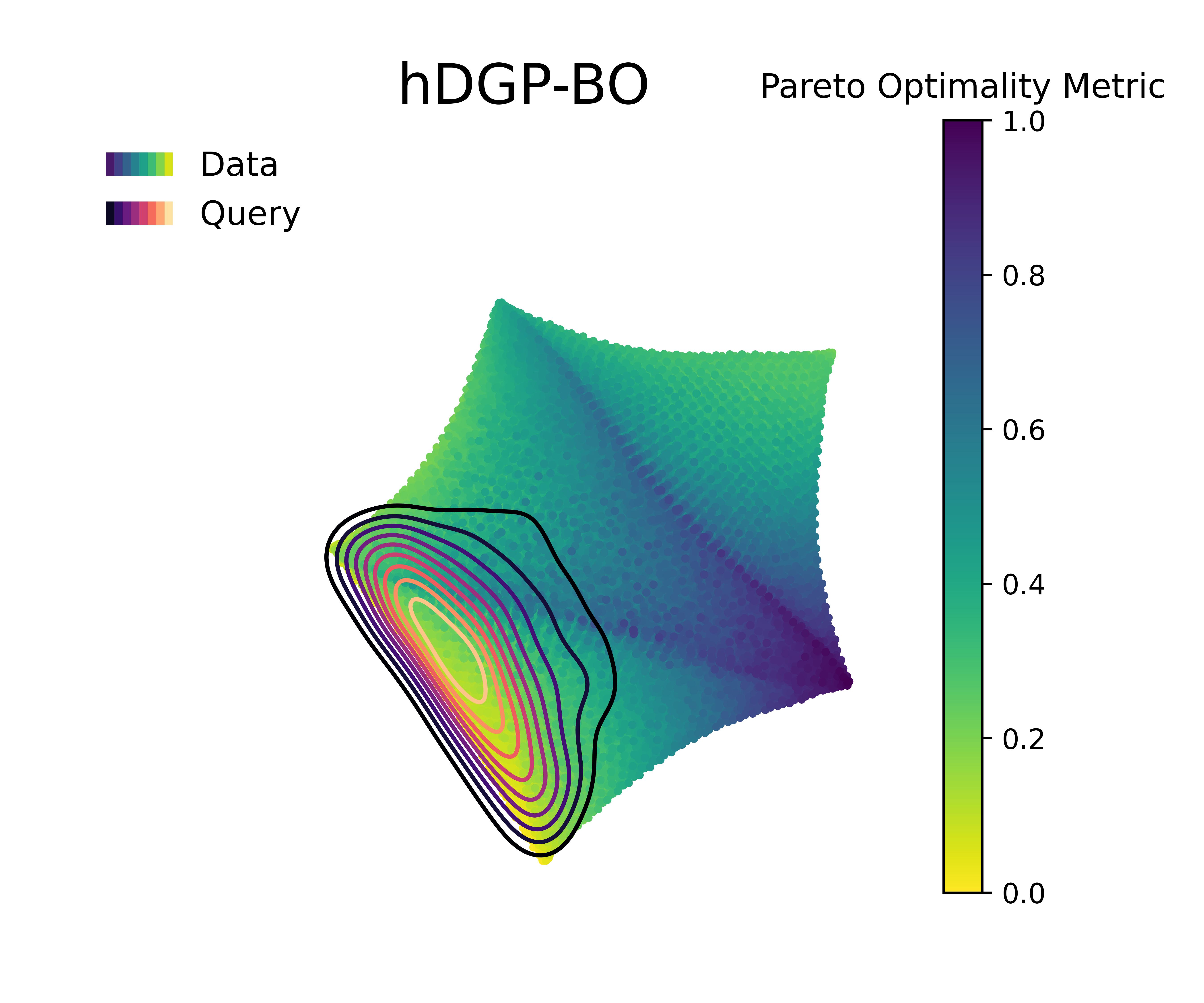}
     \caption{}
     \label{fig:queries-design-d}

 \end{subfigure}
 \caption{The initial 200 queries in design space by (a) cGP-BO, (b) MTGP-BO, (c) iDGP-BO and (d) hDGP-BO (lower Pareto optimality metric is better)}
 \label{fig:queries-design}
\end{figure}

\textbf{Figure \ref{fig:queries-objective}} \&
\textbf{Figure \ref{fig:queries-design}} visualizes the queries in the objective space and design space for the first 100 queries, and the Pareto optimality metric across the query space, respectively. The Pareto optimality metric is defined by the euclidean distance to the nearest Pareto point, scaled between 0 and 1, where 1 implies being farthest from the nearest Pareto optimal points and 0 indicate the Pareto set itself. The KDE (kernel density estimation) contounrs represent the distribution of queries made by 20 independent replications of the whole BO loop. \\ 

\textbf{Figure \ref{fig:hv-minmax}} along with \textbf{Figure \ref{fig:queries-objective}} \&
\textbf{Figure \ref{fig:queries-design}}  compare the performance and querying behavior of cGP-BO, MTGP-BO, isotopic DGP-BO (iDGP-BO), and heterotopic DGP-BO (hDGP-BO). DGP-BO and MTGP-BO models, in general, exhibit a more thorough exploration of the objective space, efficiently identifying regions with low CTE and high BM. 
The superior performance of MTGP-BO and DGP-BO can be attributed to their ability to exploit correlations between different objectives, as noted by prior works \cite{hebbal2019multi,swersky2013multi,wang2016bayesian}. 
This leads to fewer iterations to identify optimal solutions, thereby reducing computational costs and time.

A careful examination of \textbf{Figure \ref{fig:hv-minmax}} along with \textbf{Figure \ref{fig:queries-objective}} \&
\textbf{Figure \ref{fig:queries-design}} further highlights differences between iDGP-BO, hDGP-BO \& MTGP-BO. The hDGP-BO was by far performing best, showing excellent hypervolume improvements compared to others, covering more than 95\% of the actual maximum hypervolume within the first 50 iterations. In comparison, MTGP-BO (the second best model) required more than 100 iterations to achieve the same coverage. The performance bounds obtained in multiple replications of hDGP-BO was also narrower, demonstrating a lower sensitivity to initial conditions, presumably due to the additional information gained from the additional interleaved queries of BM and volume using UCB.

This interpretation is supported by the fact that iDGP-BO failed to match the performance of hDGP-BO, while sharing the same architecture. iDGP-BO was also performing worse than MTGP-BO in latter stages. 
This suggests that UCB-based uncertainty-minimization improves the performance of DGP. 
The query pattern in \textbf{Figure \ref{fig:hv-minmax}} \& \textbf{Figure \ref{fig:queries-objective}} indicates that iDGP-BO made queries roughly randomly, showing no clear sign of systematic exploration.
MTGP-BO was able to perform better than iDGP-BO and ultimately reached the hypervolume attained by hDGP-BO at latter stages. But, it was not able to achieve that as fast as hDGP-BO. \textbf{Figure  \ref{fig:hv-minmax}} \& \textbf{Figure \ref{fig:queries-objective}} indicates the querying pattern of MTGP-BO is less concentrated around the Pareto front compared to hDGP-BO, leading to lower performance.

This phenomenon may be related to the simpler linear relationship assumed between tasks by the MTGP's structure.

This more efficient exploration and exploitation of the design space by hDGP-BO corroborates findings by Hernandez-Lobato et al. (2016) \cite{hernandez2016b}, who demonstrated the effectiveness of advanced BO models in multi-objective optimization tasks through the use of acquisition functions based on information-theoretic approaches. Our obsevations indicate that the combination of an UCB heterotopic sampling strategy and of the non-linear cross-task representation lead to the superior performance of hDGP-BO.

\subsection{Multi-objective Bayesian Optimization with Auxiliary Tasks in FeCrNiCoCu Alloy Space: Maximization of TEC - Maximization of BM}

In the second optimization scenario, the objective was to maximize both the CTE and the BM. 
Finding an alloy with both high BM and high CTE is generally challenging due to the typical inverse relationship between these properties in most materials \cite{ashby2005materials}, since i) still interatomic interactions tend to oppose thermal expansion \cite{pugh1954xcii,kittel2004introduction} and an empirically-observed inverse correlation between anharmonicity (which is the driving force for CTE) and bond strength \cite{barron1999heat,fultz2010vibrational}.


    

Materials possessing both high BM and high CTE simultaneously, while relatively rare, find valuable applications in specific engineering contexts. In thermal management systems for electronics, they can provide structural stability while allowing for thermal expansion to match surrounding components. This property combination is particularly useful in electronic packaging, where managing thermal stresses is crucial \cite{chen2016thermal}. High-temperature sensors operating in extreme temperature environments also benefit from materials that maintain structural integrity while being sensitive to temperature changes. The high bulk modulus ensures the sensor's structural stability under pressure, while the high CTE allows for enhanced sensitivity to temperature variations \cite{wang2018high}. In the field of thermal energy storage, materials that can withstand high pressures and exhibit significant volume changes with temperature are desirable. These properties are particularly relevant for phase change materials and other advanced thermal storage systems, where the ability to maintain structural integrity under varying thermal conditions is critical \cite{pielichowska2014phase}. Micro-electro-mechanical systems (MEMS) and certain types of actuators also benefit from materials combining high stiffness with high thermal responsiveness. The high bulk modulus provides the necessary structural rigidity, while the high CTE allows for thermally-driven actuation. This combination is particularly valuable in the development of advanced MEMS devices and smart materials for various applications, including aerospace and robotics \cite{wu2019additively}. It's important to note that the specific requirements for "high" values of these properties can vary depending on the application context. The development of materials that optimize both properties simultaneously remains an active area of research in materials science and engineering, driving innovation in various technological fields.

\begin{figure}[H]
\centering
\includegraphics[width=1.05\linewidth]{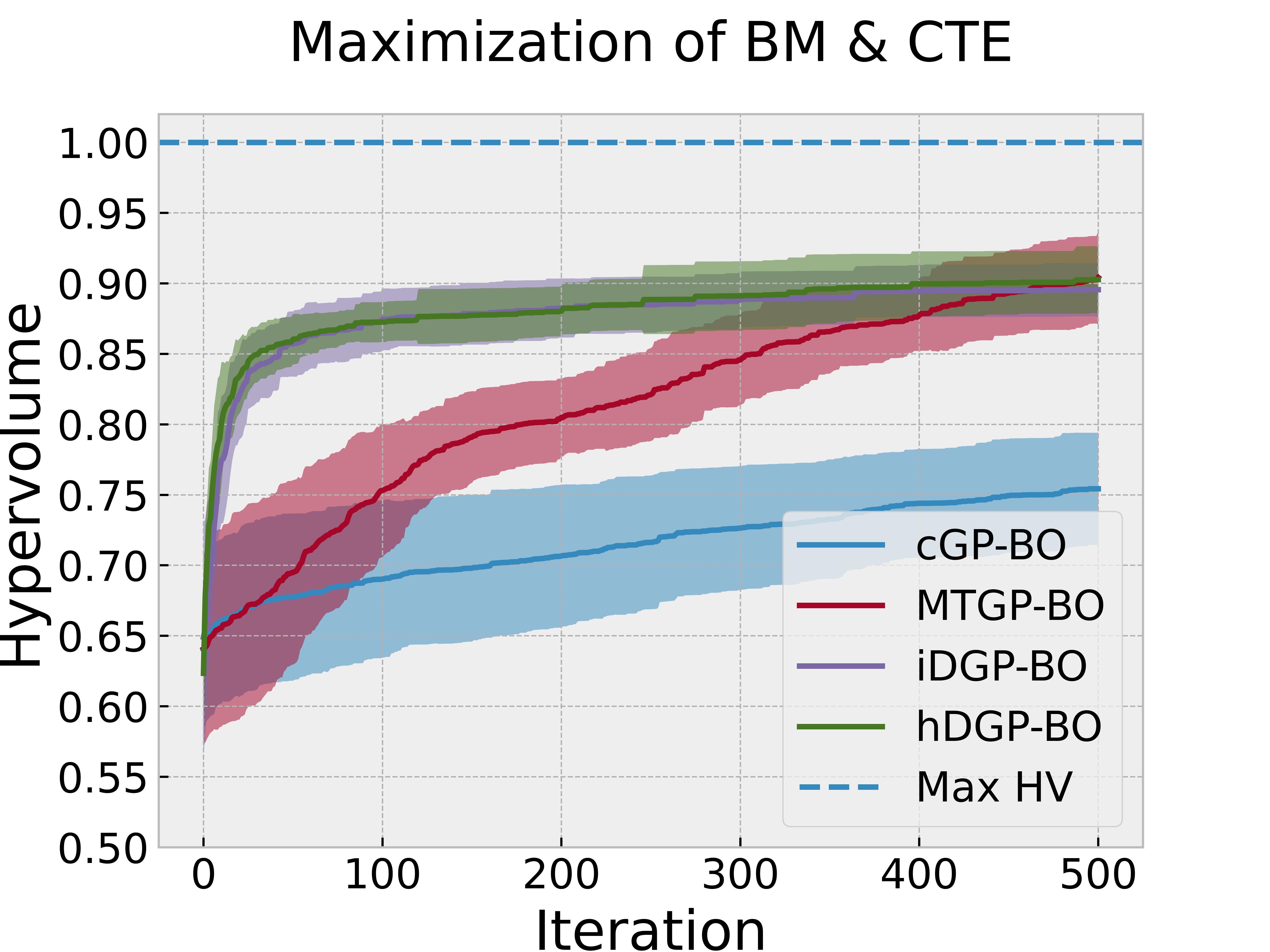}
\caption{\label{fig:hv-maxmax}The hypervolume(HV) per iteration for BO based on cGP(conventional GP), MTGP(Multi-task GP), iDGP(isotopic Deep GP) \& hDGP(heterotopic Deep GP) }
\end{figure}

\begin{figure}[H]
 \centering
 \begin{subfigure}{0.49\textwidth}
     \includegraphics[width=\textwidth]{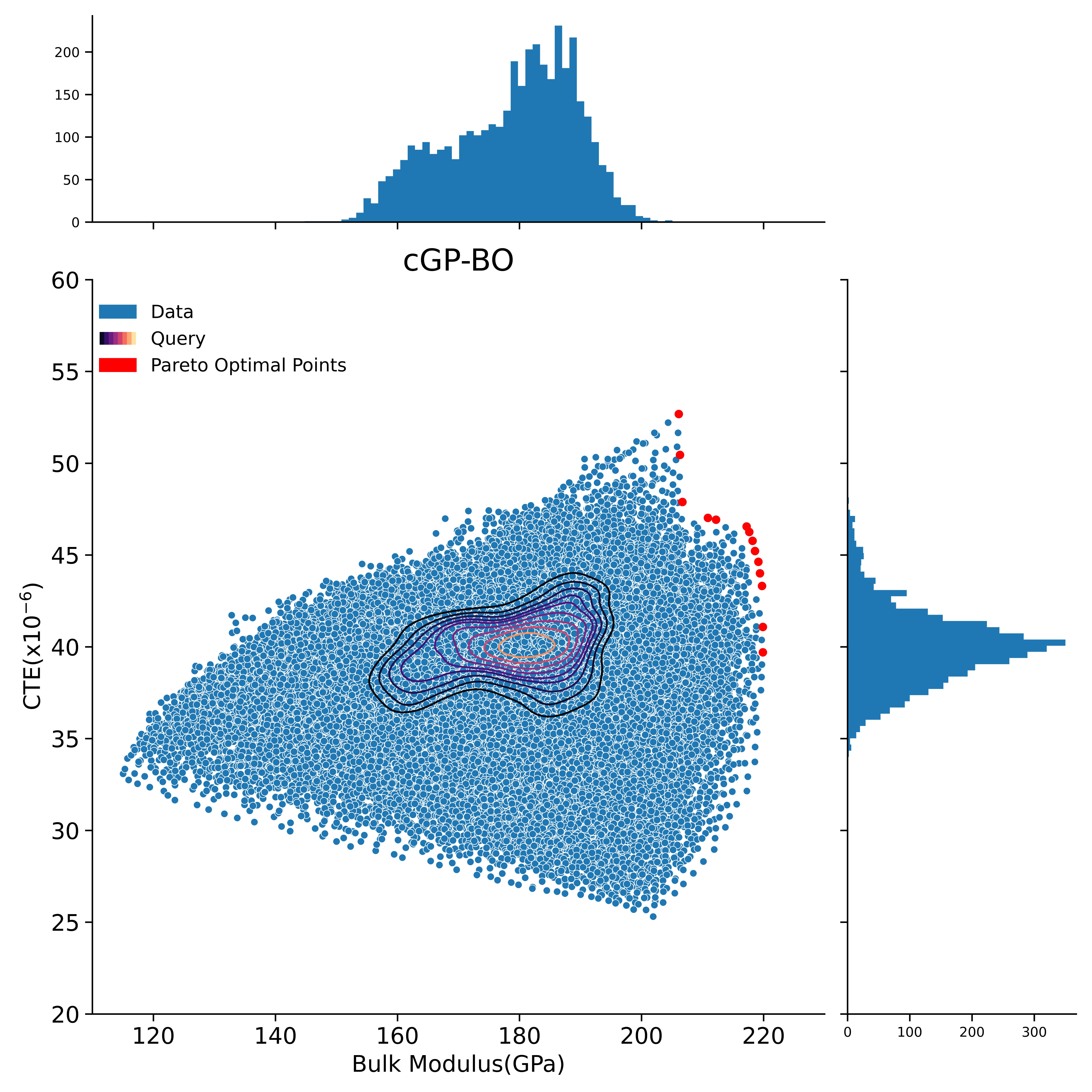}
     \caption{}
     \label{fig:queries-objective-max-a}
 \end{subfigure}
 \hfill
 \begin{subfigure}{0.49\textwidth}
     \includegraphics[width=\textwidth]{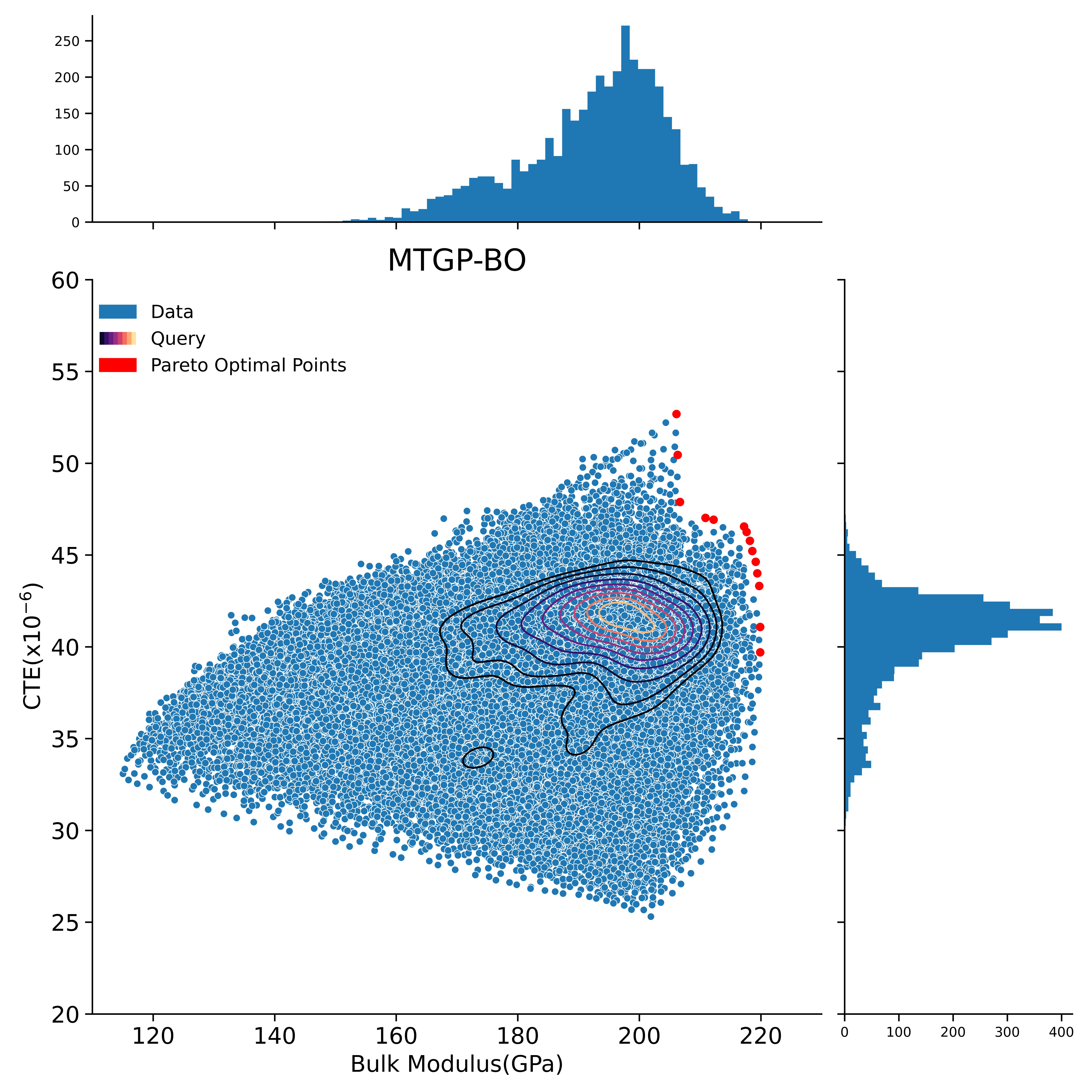}
     \caption{}
     \label{fig:queries-objective-max-b}
 \end{subfigure}
 
 \medskip
 \begin{subfigure}{0.49\textwidth}
     \includegraphics[width=\textwidth]{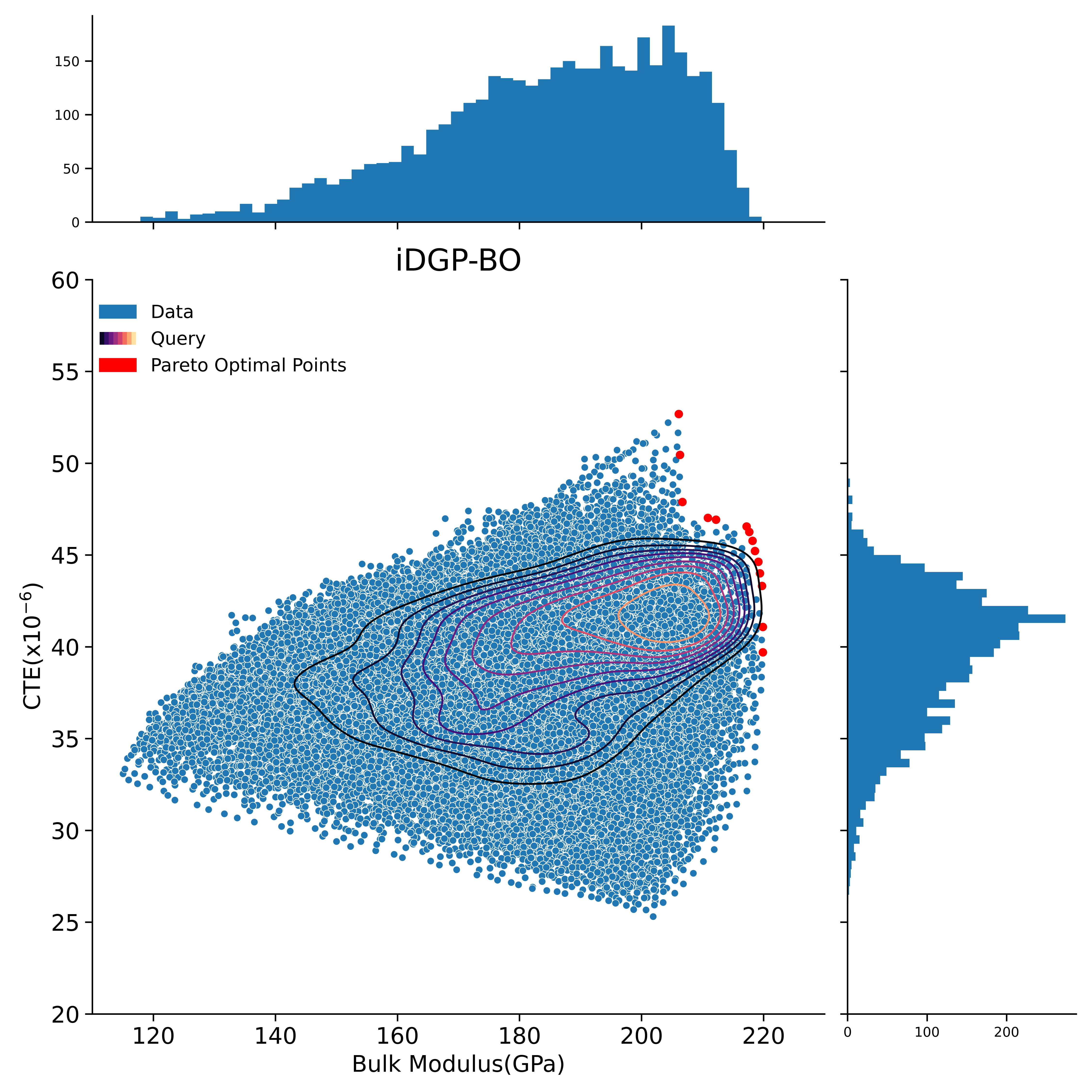}
     \caption{}
     \label{fig:queries-objective-max-c}
 \end{subfigure}
 \hfill
 \begin{subfigure}{0.49\textwidth}
     \includegraphics[width=\textwidth]{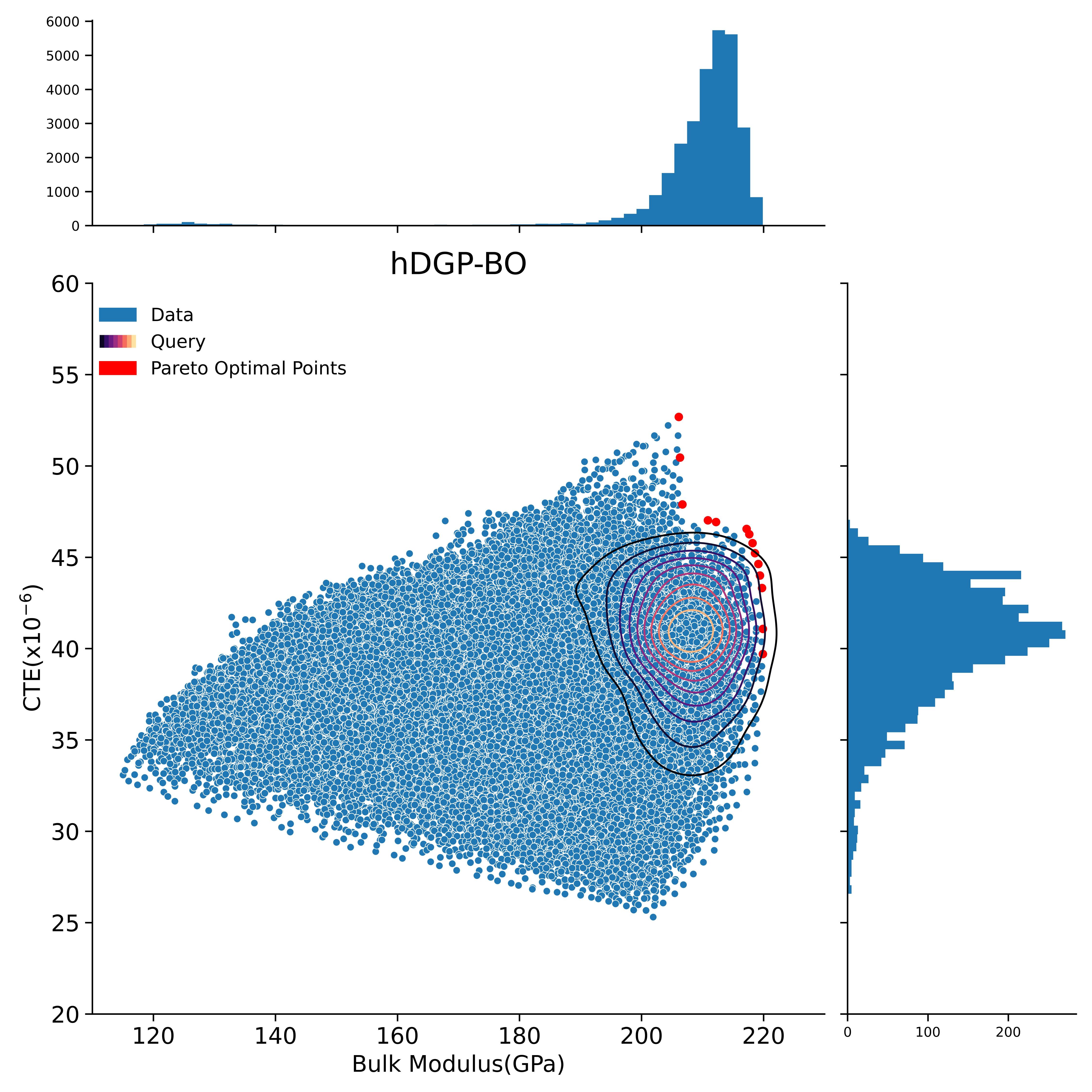}
     \caption{}
     \label{fig:queries-objective-max-d}
 \end{subfigure}

 \caption{The initial 200 queries in objective space by (a) cGP-BO, (b) MTGP-BO, (c) iDGP-BO and (d) hDGP-BO}
 \label{fig:queries-objective-max}
\end{figure}

\begin{figure}[H]
 \centering
 \begin{subfigure}{0.49\textwidth}
     \includegraphics[width=\textwidth]{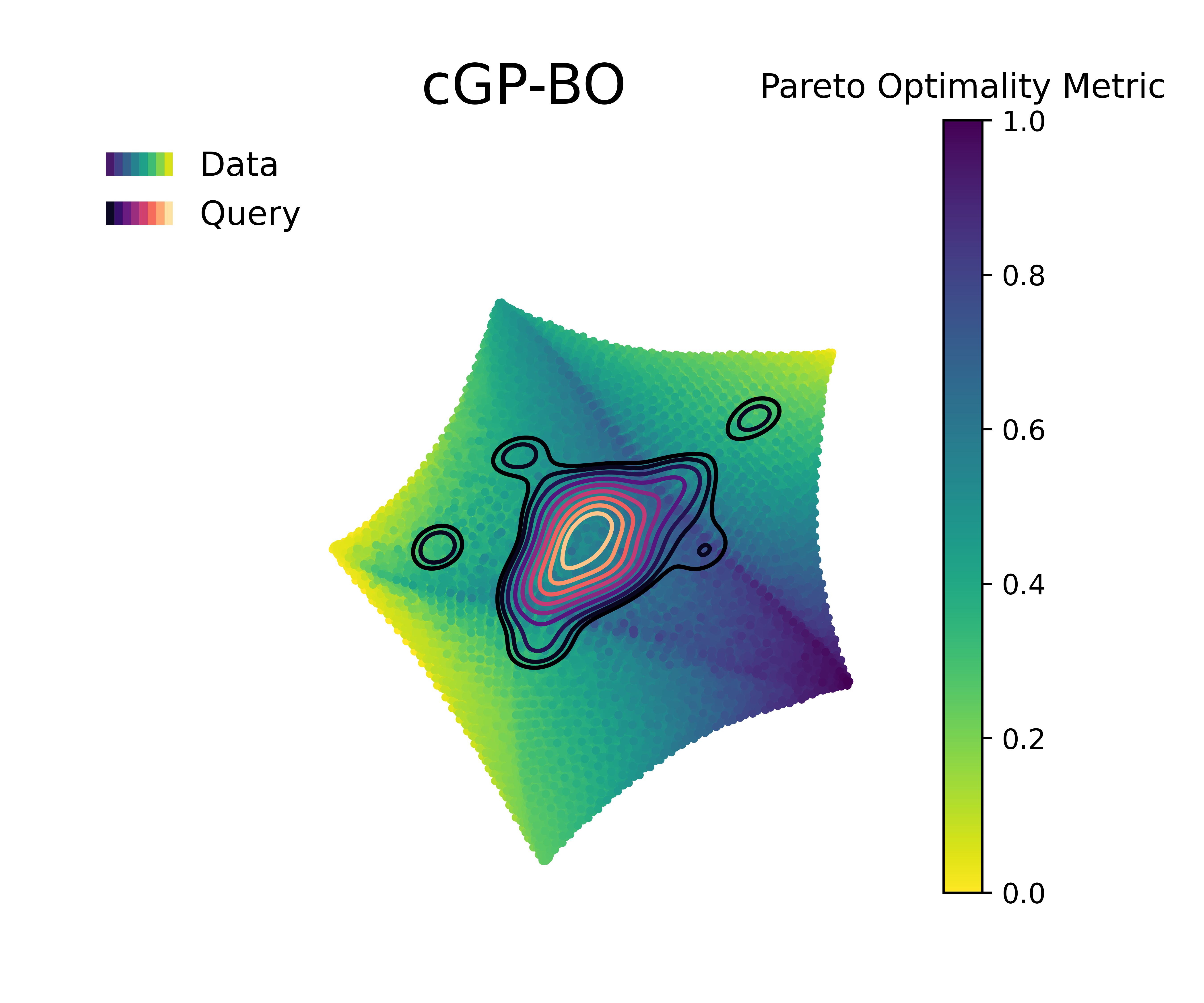}
     \caption{}
     \label{fig:queries-design-max-a}
 \end{subfigure}
 \hfill
 \begin{subfigure}{0.49\textwidth}
     \includegraphics[width=\textwidth]{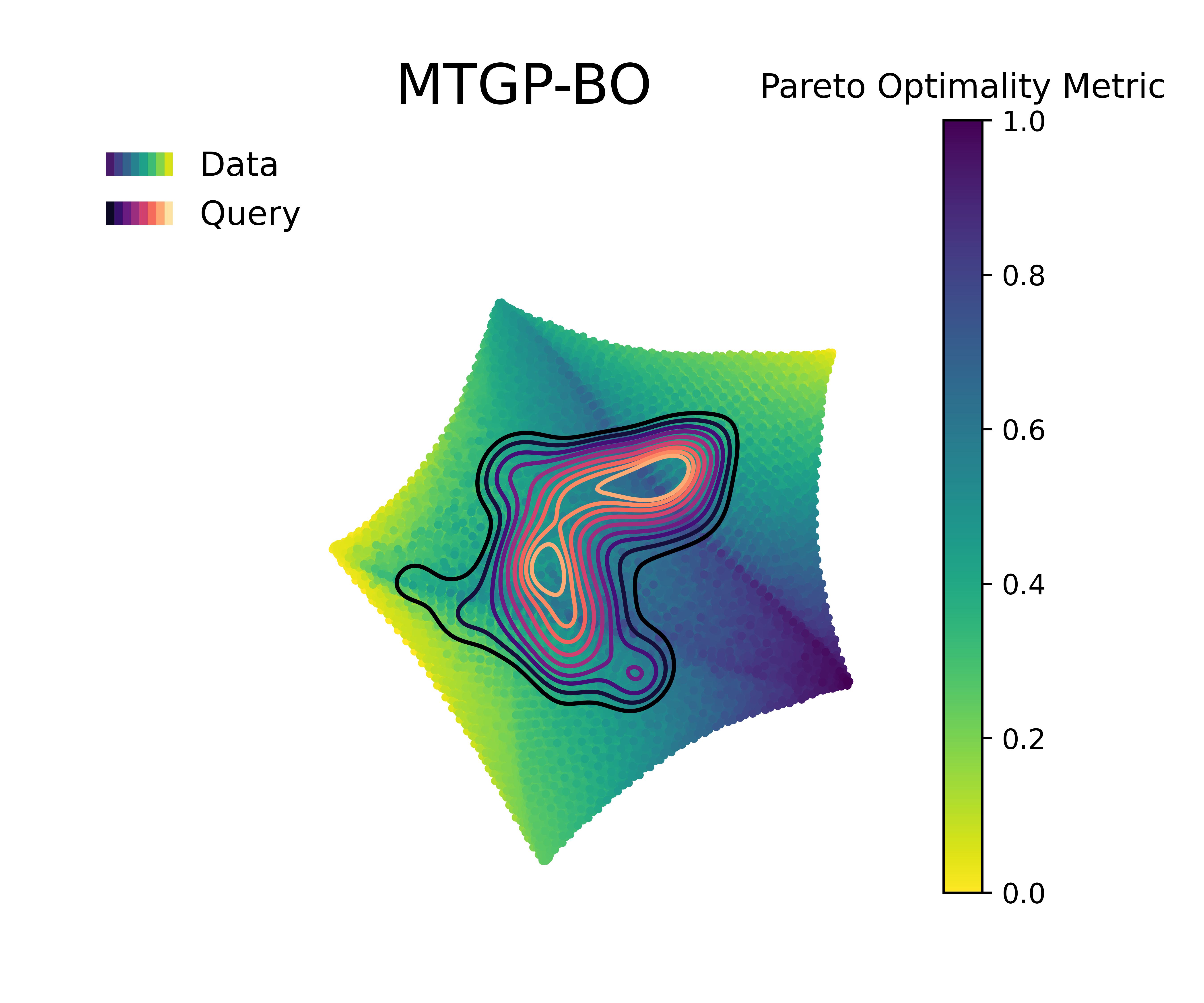}
     \caption{}
     \label{fig:queries-design-max-b}
 \end{subfigure}
 
 \medskip
 \begin{subfigure}{0.49\textwidth}
     \includegraphics[width=\textwidth]{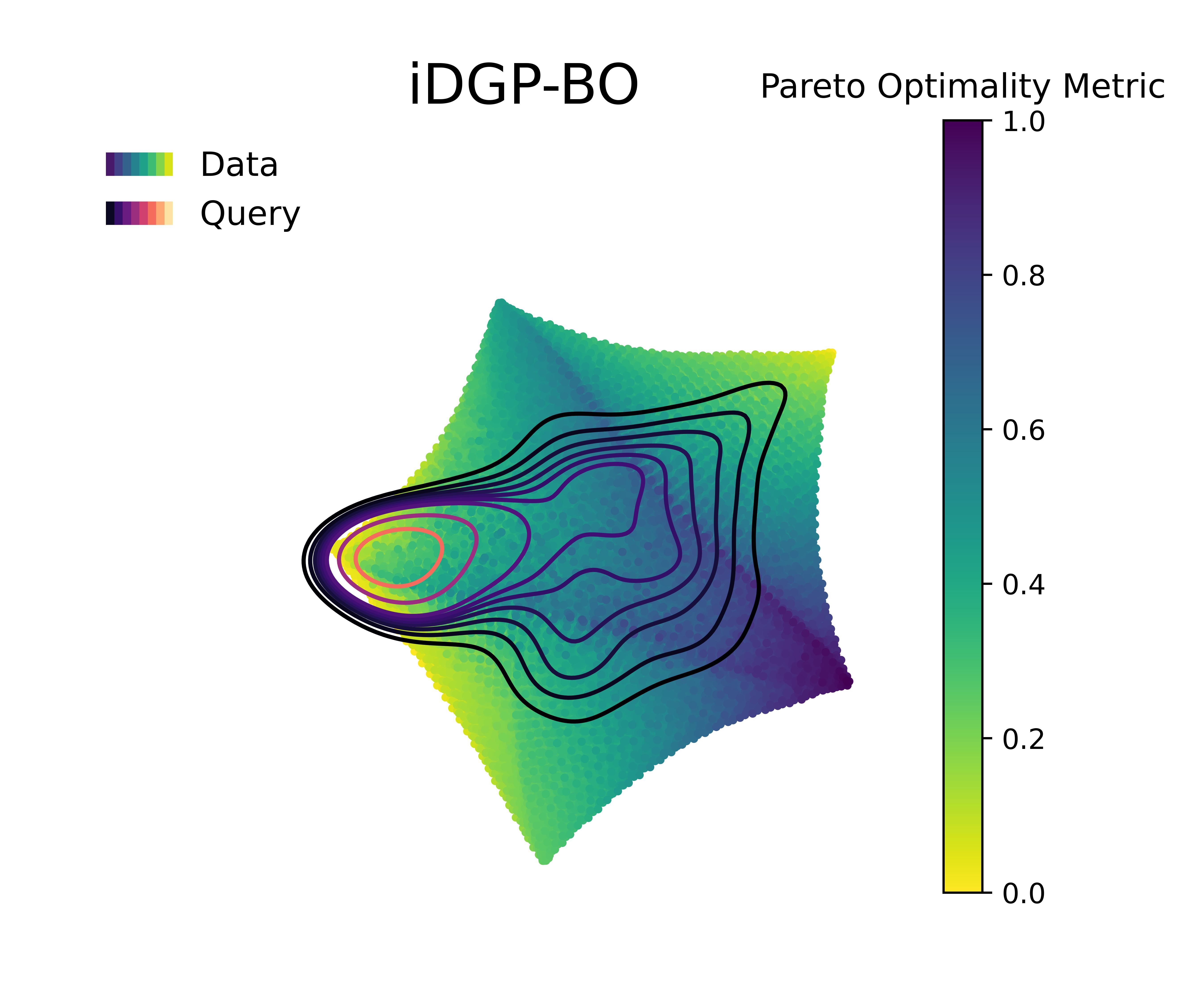}
     \caption{}
     \label{fig:queries-design-max-c}
 \end{subfigure}
 \hfill
 \begin{subfigure}{0.49\textwidth}
     \includegraphics[width=\textwidth]{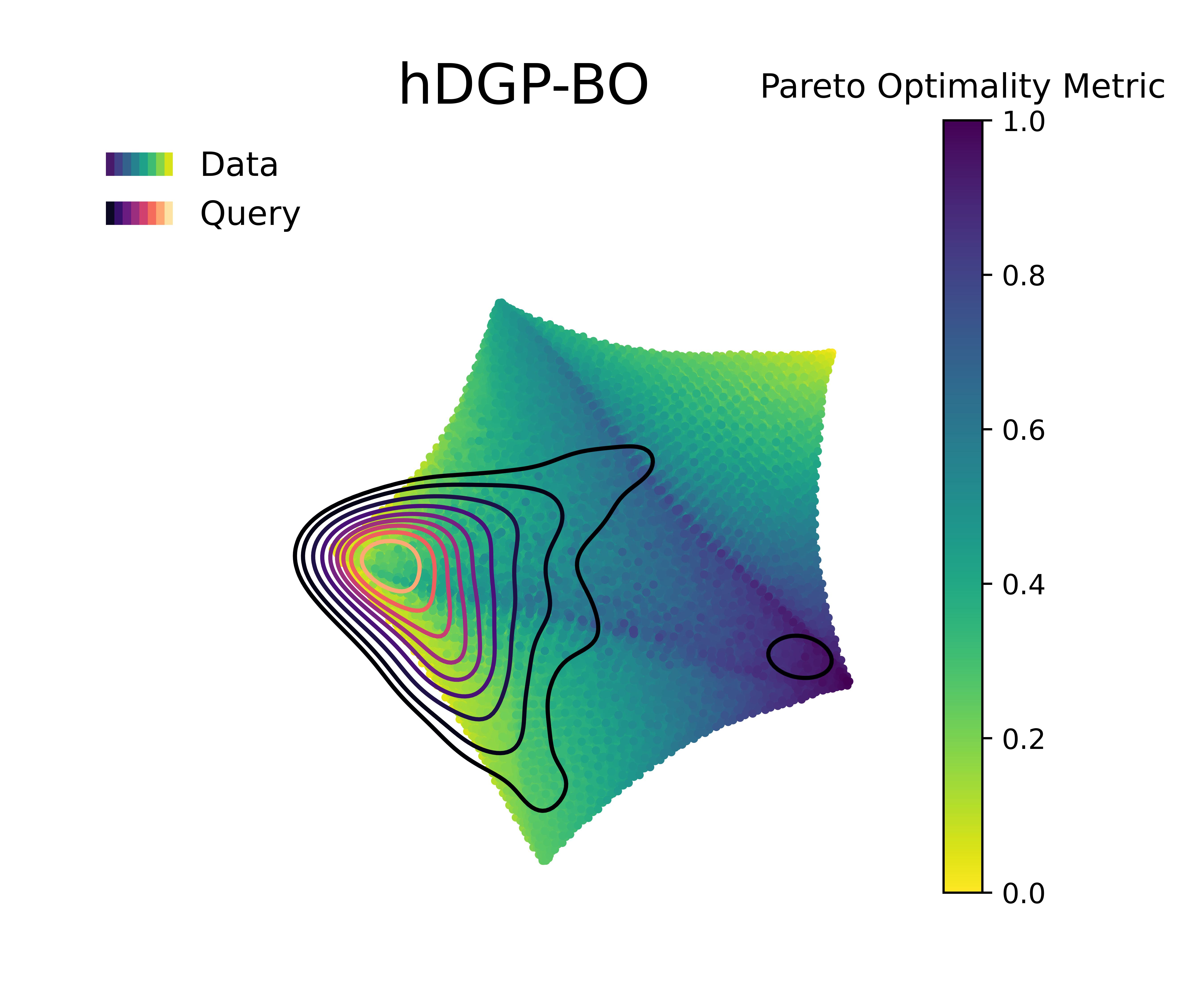}
     \caption{}
     \label{fig:queries-design-max-d}
 \end{subfigure}

 \caption{The initial 200 queries in design space by (a) cGP-BO, (b) MTGP-BO, (c) iDGP-BO and (d) hDGP-BO (lower Pareto optimality metric is better)}
 \label{fig:queries-design-max}
\end{figure}

Following the structure of the previous section, \textbf{Figure \ref{fig:hv-maxmax}} reports the evolution of the hypervolume during an iterative BO procedure based on the four different models.
\textbf{Figure \ref{fig:queries-objective-max}} \&
\textbf{Figure \ref{fig:queries-design-max}} visualizes the queries in the objective space and design space for the first 100 queries.
\textbf{Figure \ref{fig:queries-objective-max}} also displays the Pareto optimal points for reference, whereas \textbf{Figure \ref{fig:queries-design-max}} reports the corresponding Pareto optimally metric, defined as described above.

\textbf{Figure \ref{fig:hv-maxmax}} along with \textbf{Figure \ref{fig:queries-objective-max}} \&
\textbf{Figure \ref{fig:queries-design-max}} displays the difference in performance and querying behavior of cGP-BO, MTGP-BO, iDGP-BO, and hDGP-BO in this specific scenario. As in the previous problem, DGP-BO and MTGP-BO outperform cGP-BO, generating a more systemic exploration of the objective space, efficiently identifying regions with high CTE and high BM. \textbf{Figure \ref{fig:hv-maxmax}} indicates that MTGP-BO and both variants of DGP-BO (heterotopic and isotropic) outperform cGP-BO in terms of hypervolume improvement per iteration.

A closer scrutiny of \textbf{Figure \ref{fig:hv-maxmax}} along with \textbf{Figure \ref{fig:queries-objective-max}} \&
\textbf{Figure \ref{fig:queries-design-max}} reveals that, unlike the first case, iDGP-BO no longer performs worse than MTGP-BO, but instead performs better, on par with hDGP-BO. 
The difference in performance between both DGP methods on the one hand and MTGP-BO and cGP-BO on the other hand is now even wider compared to the previous scenario (c.f., \textbf{Figure \ref{fig:hv-maxmax}} vs \textbf{Figure \ref{fig:hv-minmax}} )
Both DGP methods systematically hone in to the optimal region, while the queries from MTGP-BO \& cGP-BO are more diffuse and less systematic, leading to a poor exploration of Pareto-optimal solutions, which is consistent with the poor performance exhibited in \textbf{Figure \ref{fig:hv-maxmax}}.\\

The superior exploration and search capability of hDGP-BO can be attributed to the same reasons as stated in the previous scenario. But, improvement of iDGP-BO performance over MTGP-BO along with overall decrease in performance of cGP-BO and MTGP-BO points to the difficulty of the current optimization goals. It can be inferred that, as the problem became harder, the simple linear assumption of MTGP-BO falters during earlier iterations. It demonstrated slower convergence to maximum attainable hypervolume compared to iDGP-BO and hDGP-BO. But, at latter stages, when MTGP-BO was able to explore enough data points, it slowly achieved the same level of coverage as that of DGP based BOs. So, at latter stages MTGP-BO matches the performance with that of DGP based BOs. But, during earlier iterations (within the first 50 iterations) DGP based BOs showed exceptional performance in covering close to 90\% of the maximal hypervolume, compared to MTGP-BO which achieved only 75\% coverage after 100 iterations. 
Additionally, cGP-BO performs even worse due to the enhanced difficulty of this problem. The difficulty faced by MTGP-BO and cGP-BO is also reflected in the querying pattern(
\textbf{Figure \ref{fig:queries-objective-max}} \&
\textbf{Figure \ref{fig:queries-design-max}}).
In both of the cases, the hDGP-BO proposed in this paper is the most robust. 

\subsection{Mutual Information Analysis}

To further analyze the differences in performance among multiple variants of BO, an analysis based on mutual information(MI) is carried out. The prediction of the four BOs were recorded in each iteration at 5000 points(sampled through dirchlet distribution) for each tasks. Then, the mutual information between the two objectives were calculated from the predicted 5000 points. This was done at each iteration.\\
\begin{figure}[H]
 \centering
 \begin{subfigure}{0.49\textwidth}
     \includegraphics[width=\textwidth]{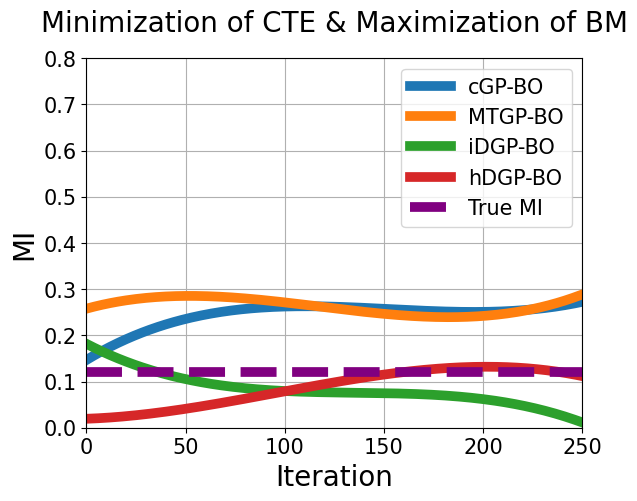}
     \caption{}
     \label{fig:MI-01}
 \end{subfigure}
 \hfill
 \begin{subfigure}{0.49\textwidth}
     \includegraphics[width=\textwidth]{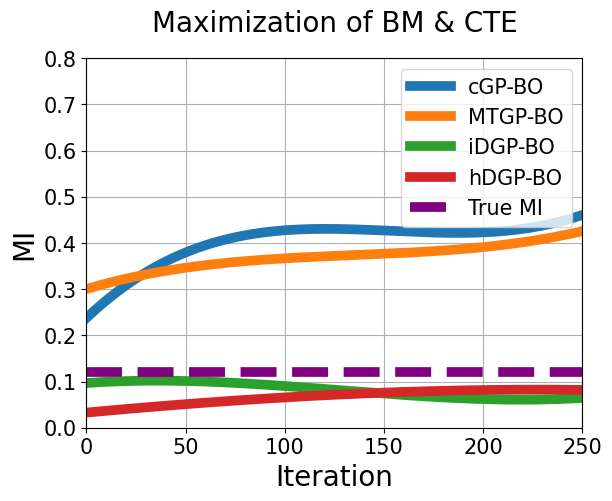}
     \caption{}
     \label{fig:MI-11}
 \end{subfigure}
 \caption{MI between predicted BM and CTE at each iteration for cGP-BO, MTGP-BO, iDGP-BO and hDGP-BO, along with the true MI calculated from the ground truth}
 \label{fig:MI-itr}
\end{figure}
In the first scenario, \textbf{Figure \ref{fig:MI-01}} shows that, only the predicted MI of hDGP-BO gets close to the true MI. Whereas, iDGP-BO struggles to achieve any significant MI due to its unsystematic query pattern(\textbf{Figure \ref{fig:queries-design-c}}). MTGP-BO display high predicted MI due to the fact that they assume linear correlation between tasks. High MI translates into simpler and linear relationship. The cGP-BO also shows high MI without having correlated kernel.It's because the different tasks start out with same prior, making the tasks highly correlated at unexplored regions. The agreement between predicted and true MI translates into the performance difference(\textbf{Figure \ref{fig:hv-minmax}}). The BO having the best agreement(hDGP-BO) shows the best performance.

The second scenario(\textbf{Figure \ref{fig:MI-11}}) reaffirms the hypothesis "ability to achieve correct predicted MI translates into better optimization performance". Here, it is observed that iDGP-BO is able to capture the correct predicted MI(unlike first scenario). As a result, the iDGP-BO also performs better(\textbf{Figure \ref{fig:hv-maxmax}}), unlike the first case. Moreover, the predicted MI from cGP-BO and MTGP-BO shows significant deviation from true MI in this second case. As a result, their optimization performance also deteriorates compared to the first scenario(\textbf{Figure \ref{fig:hv-maxmax}}).

 In this work, predicted MI per iteration has been  demonstrated to be useful in quantifying performance for optimization problems having multiple correlated tasks. Further research is warranted to investigate the use of predicted MI in different multi-objective optimization problems having multiple correlated tasks. It can be proposed as a metric of the degree to which a model can exploit correlations.

\subsection{Addition of Random Gaussian Field Noise}

In this subsection, we analyze the impact of adding random Gaussian field noise to the Bayesian optimization process. The addition of noise is essential for evaluating the robustness and stability of the optimization algorithms under real-world conditions, where measurements and simulations are often subject to various sources of uncertainty \cite{williams2006,rasmussen2003gaussian}.

Gaussian field noise is characterized by its mean and covariance structure, which can be defined as:

\[
\epsilon(\mathbf{x}) \sim \mathcal{N}(0, k_{\epsilon}(\mathbf{x}, \mathbf{x}'))
\]

where \(k_{\epsilon}(\mathbf{x}, \mathbf{x}')\) is the covariance function of the noise, typically chosen to be a squared exponential kernel:

\[
k_{\epsilon}(\mathbf{x}, \mathbf{x}') = \sigma_{\epsilon}^2 \exp\left( -\frac{\|\mathbf{x} - \mathbf{x}'\|^2}{2 l_{\epsilon}^2} \right)
\]

Here, \(\sigma_{\epsilon}^2\) represents the noise variance, and \(l_{\epsilon}\) is the length scale of the noise \cite{rasmussen2003gaussian}. In this work, The lengthscale is chosen to be 0.20. The variance is selected such that, the third standard deviation does not exceed more than half of the spread for each tasks. As a result, CTE and BM have variances of 20 and 100 respectively. This implies uncertainty in CTE data having 4.5 $\times$$10^{-6}$ K$^{-1}$ standard deviation and in BM having 10 GPa standard deviation.   

To assess the influence of this noise on the optimization performance, we conducted experiments by adding random Gaussian field noise to the objective functions during the Bayesian optimization process. The performance of the optimization algorithms was evaluated both before and after the addition of noise \cite{hermans2021overfitting}.
\begin{figure}[H]
 \centering
 \begin{subfigure}{0.49\textwidth}
     \includegraphics[width=\textwidth]{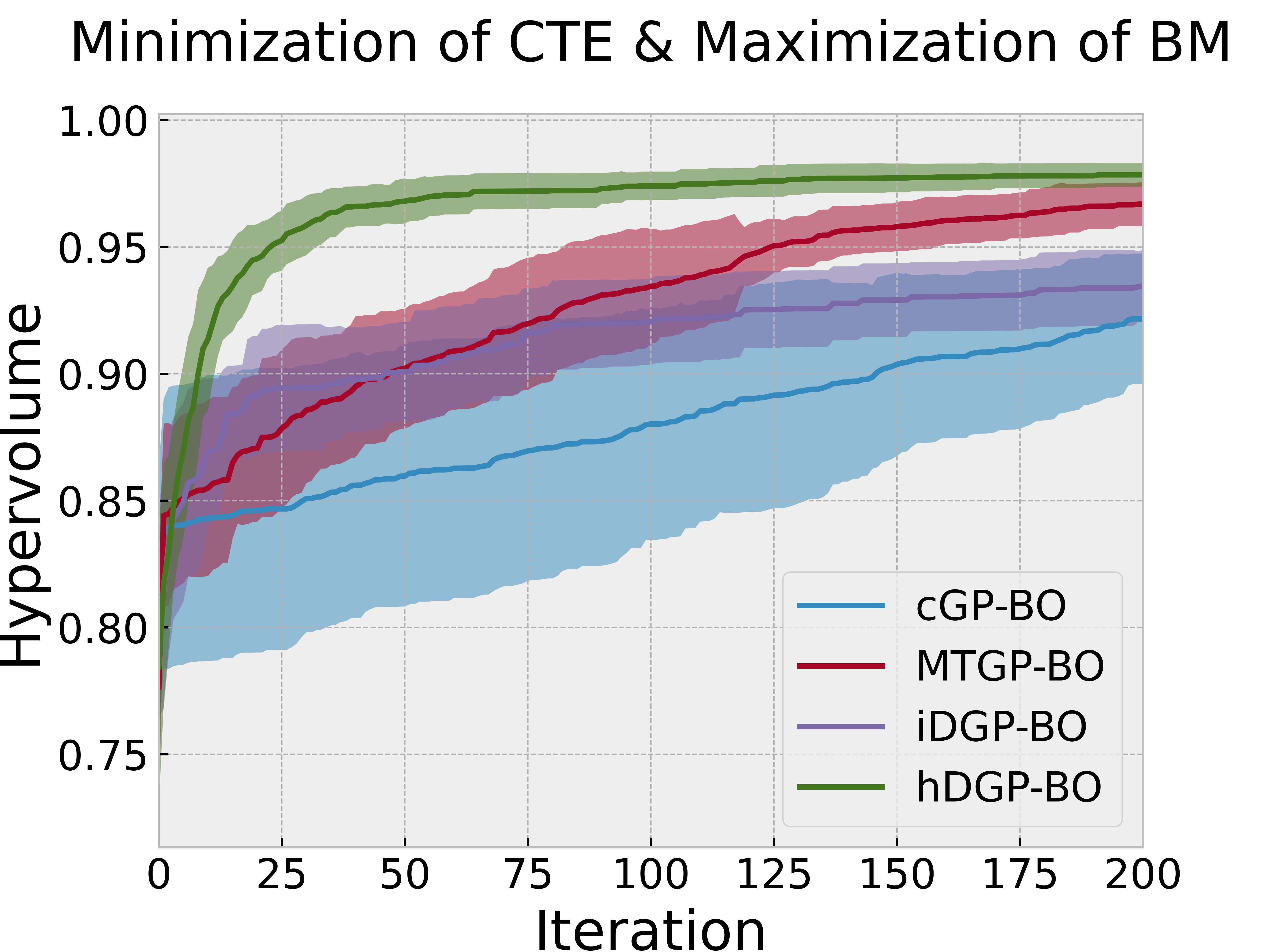}
     \caption{Performance before addition of Random Gaussian Field noise}
     \label{fig:new-minmax-200}
 \end{subfigure}
 \hfill
 \begin{subfigure}{0.49\textwidth}
     \includegraphics[width=\textwidth]{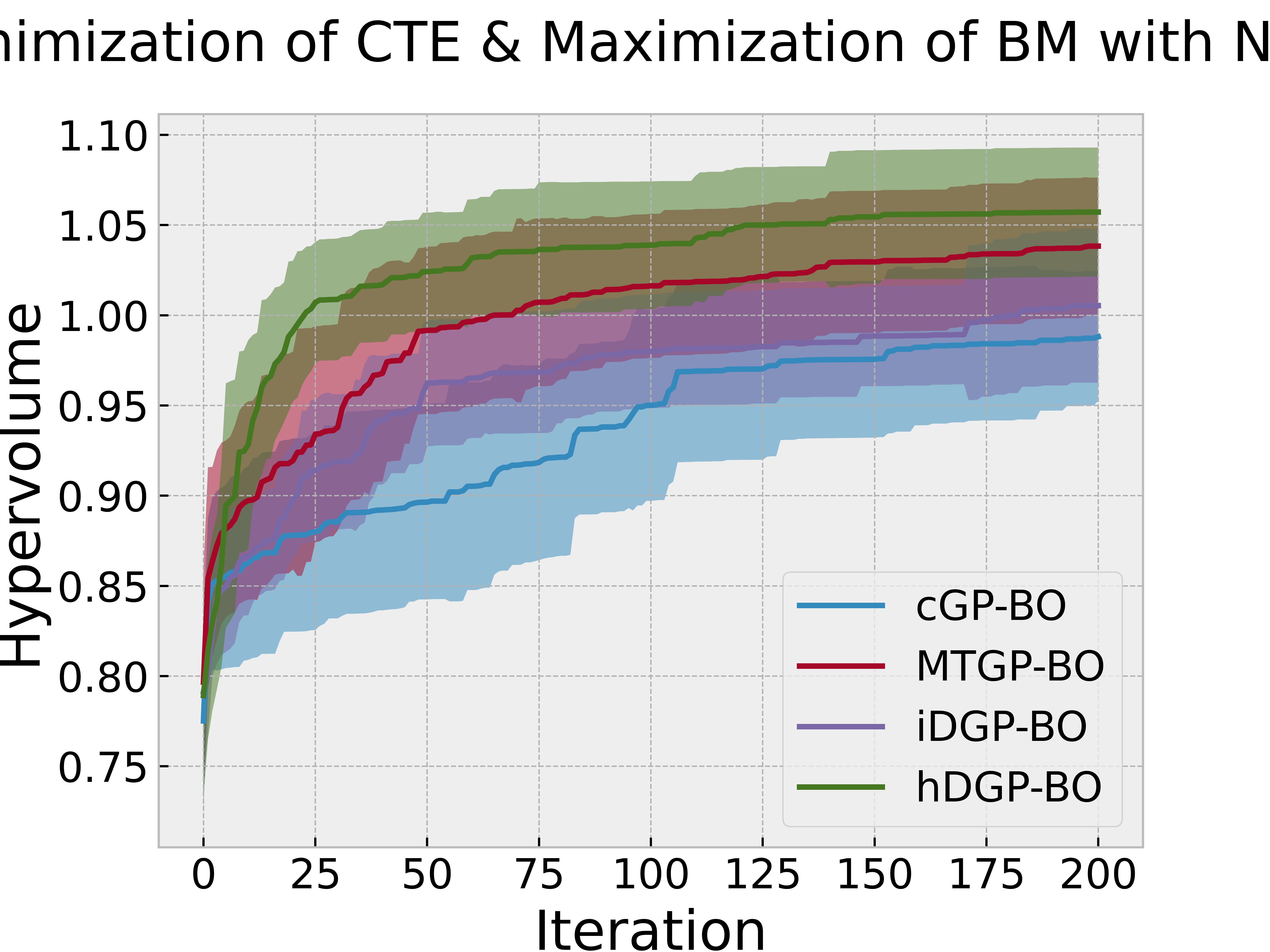}
     \caption{Performance after addition of Random Gaussian Field noise}
     \label{fig:new-minmax-grf-200}
 \end{subfigure}
 \caption{Hypervolume improvement per iteration for the minimization of TEC and maximization of BM before and after the addition of random Gaussian field noise.}
 \label{fig:hv-minmax-noise}
\end{figure}

\begin{figure}[H]
 \centering
 \begin{subfigure}{0.49\textwidth}
     \includegraphics[width=\textwidth]{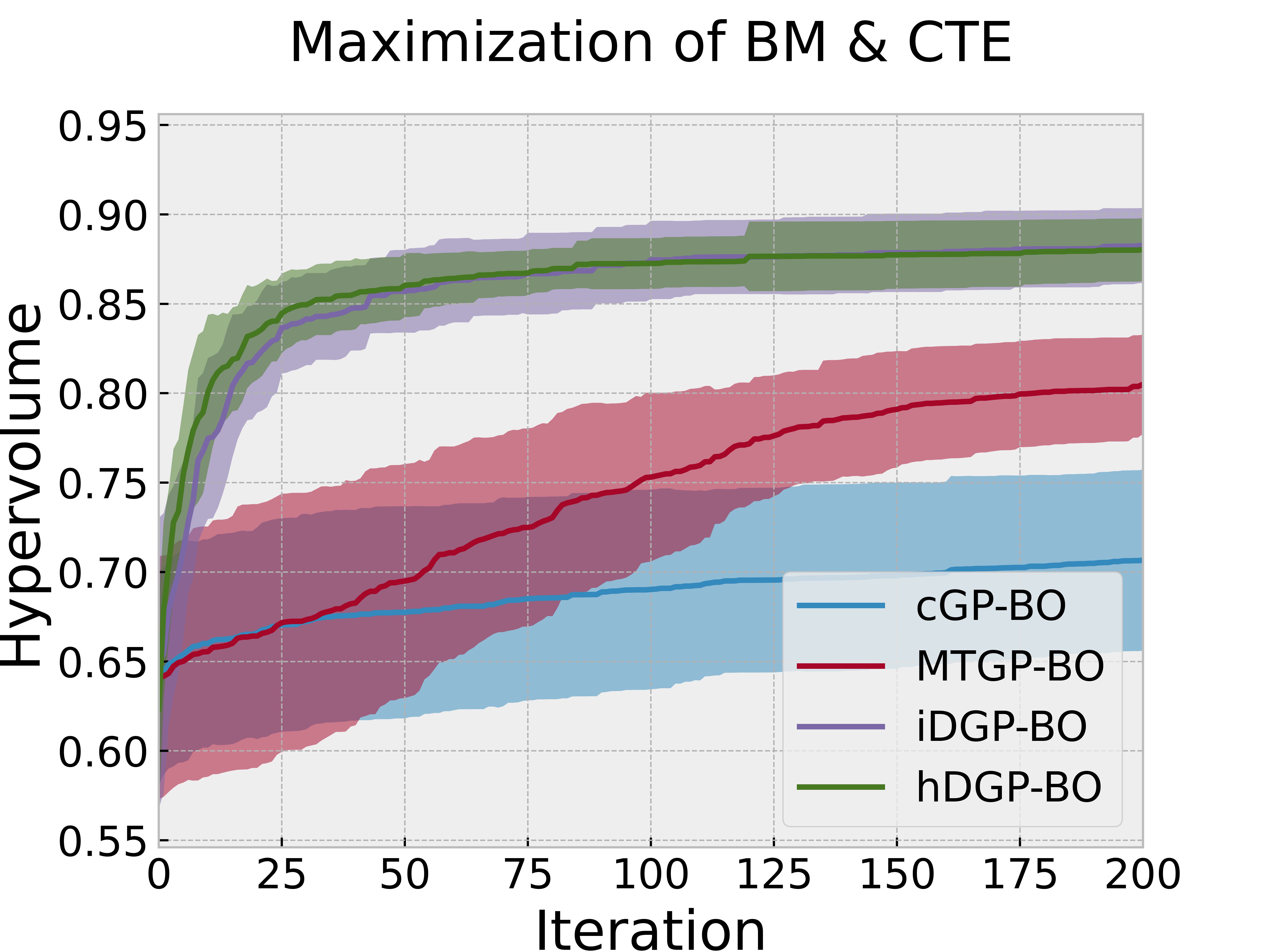}
     \caption{Performance before addition of Random Gaussian Field noise}
     \label{fig:new-maxmax-200}
 \end{subfigure}
 \hfill
 \begin{subfigure}{0.49\textwidth}
     \includegraphics[width=\textwidth]{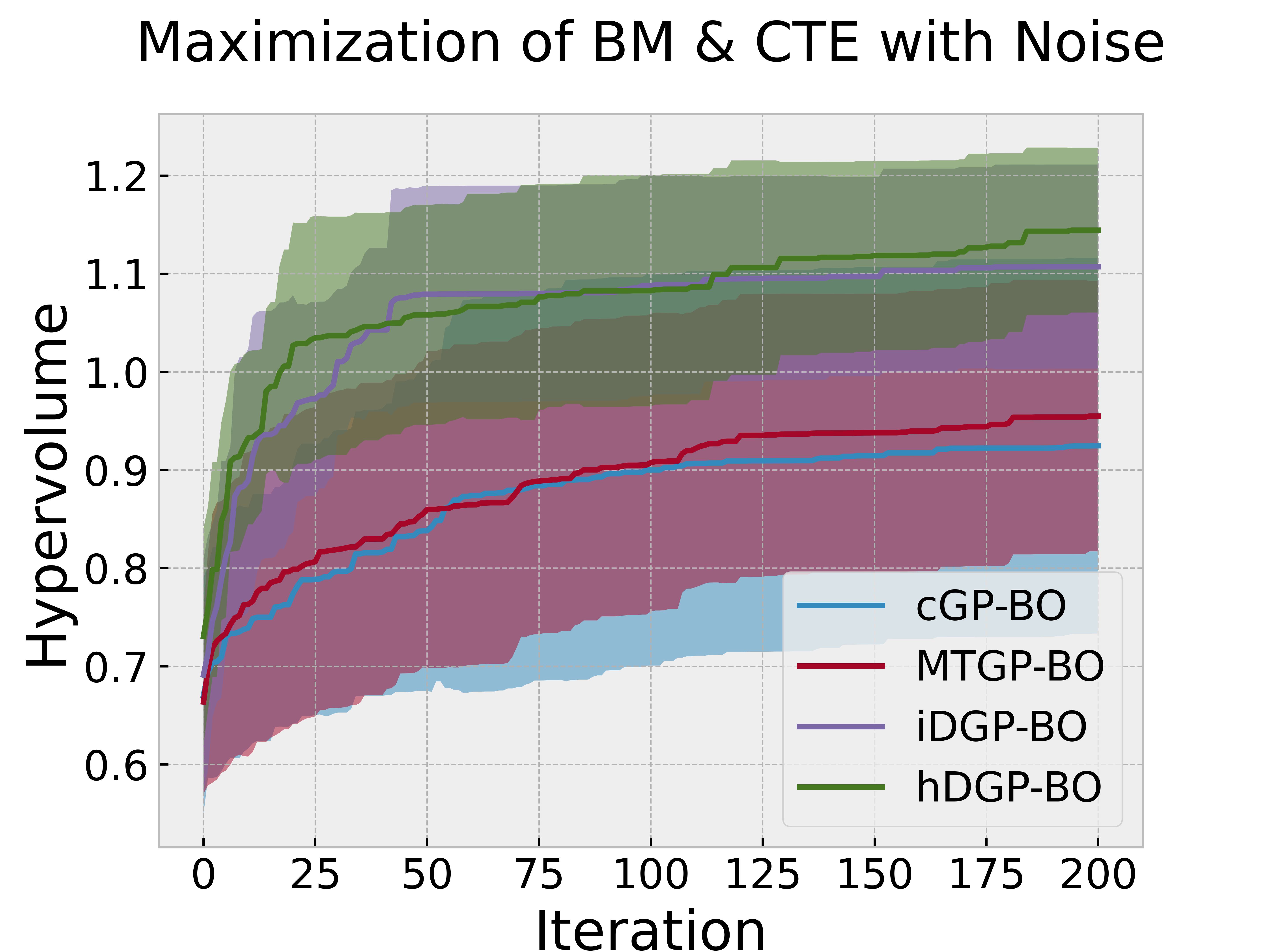}
     \caption{Performance after addition of Random Gaussian Field noise}
     \label{fig:new-maxmax-grf-200}
 \end{subfigure}
 \caption{Hypervolume improvement per iteration for the maximization of TEC and BM before and after the addition of random Gaussian field noise.}
 \label{fig:hv-maxmax-noise}
\end{figure}

\textbf{Figure \ref{fig:hv-minmax-noise}} and \textbf{Figure \ref{fig:hv-maxmax-noise}} illustrate the hypervolume improvement per iteration for the two optimization scenarios: minimization of TEC with maximization of BM, and maximization of both TEC and BM, respectively. The figures compare the performance of the optimization algorithms before and after the addition of random Gaussian field noise.

The results indicate that the addition of noise generally leads to a decrease in the optimization performance advantages between the models. The DGP-BO models still show better performance compared to the cGP-BO \& MTGP-BO model. But, the performance advantages of DGP-BOs decrease. The inherent uncertainty between multiple replications also increases. This happens due to the fact that, addition of gaussian field noise obscures the complex non-linear relationship between tasks. Thus, it becomes harder for DGPs or MTGPs to take advantage of the inter-relationship between tasks. This further proves that the DGPs and MTGPs performs better then cGP-BO by exploiting correlations. Otherwise, obscuring the correlation through noise addition would not have affected the performance difference between them.  

The hDGP-BO still demonstrates robustness to obscured correlation between tasks. This robustness can be attributed to the sophisticated kernel structure and uncertainty reduction steps using UCB.
In the context of HEA design, the presence of noise is a common challenge due to the complexity of the systems and the limitations of experimental and computational methods \cite{senkov2018development}. The robustness of hDGP-BO to noise makes them particularly suitable for handling the uncertainties inherent in HEA property data, ensuring more reliable optimization results.
The incorporation of random Gaussian field noise in the optimization process serves as a valuable test for evaluating the robustness of Bayesian optimization algorithms. The findings underscore the advantages of using advanced models hDGP-BO in scenarios with inherent uncertainties, which are common in real-world materials discovery tasks \cite{chen2016practical}.


\section{Conclusions}

Our study demonstrates that both DGP and MTGP variants of BO have significant advantages in high-throughput materials discovery involving multiple objectives. Moreover, the novel hDGP-BO in particular(proposed in this study) is the most robust and efficient in all of the cases. Utilizing hDGP-BO, we can accelerate the discovery of materials with specific properties by exploiting correlations between auxiliary properties. The superior performance of these models is attributed to their ability to capture complex, non-linear relationships, exploit correlations between objectives, and adaptively learn the inter-dependencies in the data. MTGP-BO has also demonstrated superior performance compared to iDGP-BO in specific cases, proving its importance. The robustness of hDGP-BO and MTGP-BO to random Gaussian field noise further highlights their suitability for real-world materials discovery tasks, where uncertainties are inherent in experimental and computational data \cite{chen2016practicalb,rasmussen2003gaussian}.

The correlation analysis presented in this study underscores the importance of considering property inter-dependencies in the optimization of HEAs and validates the effectiveness of  hDGP-BO or MTGP-BO in leveraging these relationships to accelerate materials discovery \cite{wang2020b}. By exploiting the correlations between thermal, mechanical, and structural properties, these advanced models can guide the search more effectively towards optimal compositions \cite{liu2018b}.

Future work includes parallelizing hDGP-BO and MTGP-BO and developing acquisition functions tailored to these models to further enhance performance. Additionally, integrating physics-based models and domain knowledge into the optimization framework can provide a more comprehensive understanding of the underlying mechanisms governing the behavior of HEAs and guide the discovery of novel compositions with targeted properties.

In conclusion, this study demonstrates the potential of advanced Bayesian optimization models like MTGP-BO and hDGP-BO in accelerating the discovery of high-performance HEAs. By exploiting property correlations, adapting to complex relationships, and robustly handling uncertainties, these models can significantly streamline the materials discovery process. The insights gained from this research can be extended to other materials systems and optimization challenges, paving the way for more efficient and targeted exploration of vast design spaces.


\bibliographystyle{elsarticle-num}

\bibliography{main.bib}

\end{document}